\documentclass{article}

\usepackage{arxiv}

\usepackage[utf8]{inputenc} 
\usepackage[T1]{fontenc}    
\usepackage{hyperref}       
\usepackage{url}            
\usepackage{booktabs}       
\usepackage{amsfonts}       
\usepackage{nicefrac}       
\usepackage{microtype}      
\usepackage{lipsum}
\usepackage{graphicx}
\usepackage{subfigure}
\usepackage{booktabs}
\usepackage{amsmath}
\usepackage{amssymb}
\usepackage{mathtools}
\usepackage{amsthm}
\usepackage{diagbox}
\usepackage{authblk}
\usepackage[english]{babel}

\usepackage{wrapfig}

\usepackage{multirow}
\usepackage{multicol}
\usepackage{bm}
\title{CTDS: Centralized Teacher with Decentralized Student for Multi-Agent Reinforcement Learning}

\author[1]{Jian Zhao}
\author[1]{Xunhan Hu}
\author[1]{Mingyu Yang}
\author[1]{Wengang Zhou}
\author[2]{Jiangcheng Zhu}
\author[1]{Houqiang Li}
\affil[1]{University of Science and Technology of China}
\affil[2]{Huawei Cloud}

\begin{document}
\maketitle
\def\thefootnote{*}\footnotetext{These authors contributed equally to this work}
\begin{abstract}
Due to the partial observability and communication constraints in many multi-agent reinforcement learning (MARL) tasks, centralized training with decentralized execution (CTDE) has become one of the most widely used MARL paradigms.
In CTDE, centralized information is dedicated to learning the allocation of the team reward with a mixing network, while the learning of individual Q-values is usually based on local observations.
 The insufficient utility of global observation will degrade performance in challenging environments.
  To this end, this work proposes a novel Centralized Teacher with Decentralized Student (CTDS) framework, which consists of a teacher model and a student model. Specifically, the teacher model allocates the team reward by learning individual Q-values conditioned on global observation, while  the student model utilizes the partial observations to approximate the Q-values estimated by the teacher model.
  In this way, CTDS balances the full utilization of global observation during training and the feasibility of decentralized execution for online inference.
  Our CTDS framework is generic which is ready to be applied upon existing CTDE methods to boost their performance. 
  We conduct experiments on a challenging set of StarCraft II micromanagement tasks to test the effectiveness of our method and the results show that CTDS outperforms the existing value-based MARL methods.
\end{abstract}


\section{Introduction}\label{sec:introduction}
Cooperative multi-agent reinforcement learning (MARL) has been demonstrating significant applicability in a variety of domains, such as autonomous cars~\cite{cao2012overview}, sensor networks~\cite{zhang2011coordinated}, robot swarms~\cite{huttenrauch2017guided,busoniu2008a}.
To train an MARL system, a group of agents coordinate with each other to learn action-values conditioned on state information, and jointly optimize a single reward signal, $i.e.$, the team reward, accumulated~\cite{sunehag2017value,hernandez-leal2019a}.

A naive solution to MARL is to convert a cooperative multi-agent problem into a single-agent RL problem by taking the joint state/action space of all the agents as the state/action space of one virtual agent~\cite{openai2019dota, ye2020mastering,foerster2016learning}, which is usually viewed as the  paradigm  of centralized training with centralized execution (CTCE).
Despite its effectiveness, cooperative MARL with centralized execution encounters a major challenge of scalability, \emph{i.e.}, the joint state-action space grows exponentially as the number of agents increases. 
Besides, in many real-world settings, due to observation and communication constraints, it becomes impractical to make centralized execution \cite{sunehag2017value,rashid2018qmix}.
Another alternative approach is the  paradigm  of decentralized training with decentralized execution (DTDE), which trains independent learners to optimize for the team reward.
However, it is difficult to design effective individual reward functions for different agents when only the team reward is available.
In addition, decentralized training neglects the coordination among agents, which is essential for multi-agent systems.
Compared to decentralized training, training the agents in a centralized fashion allows the access to the global information and removes inter-agent communication constraints, which is beneficial to capture the coordination pattern and allocate the team reward.

Considering the necessity of decentralized policies and the benefits of centralized training in MARL, the paradigm of centralized training with decentralized execution (CTDE)~\cite{kraemer2016multi, oliehoek2008optimal} has gained increasing attention in the last few years.
In this paradigm, the policy of each agent is trained with global context in a centralized way and executed only based on local histories in a decentralized way. 
There are two non-trivial issues in CTDE: credit assignment and partial observation.  Credit assignment refers to allocating the team
reward to each agent by leveraging the global state information
during the centralized training, while partial observation indicates the fact that each agent can only access a proportion of the global state during inference.


To address the above two issues, many CTDE learning approaches ~\cite{sunehag2017value,lowe2017multi,rashid2018qmix,foerster2018counterfactual,son2019qtran,yang2020qatten,wang2020qplex}  have been proposed recently, among which value based MARL algorithms~\cite{sunehag2017value,rashid2018qmix,son2019qtran,yang2020qatten,wang2020qplex} have shown state-of-the-art performance on challenging tasks, e.g., unit micromanagement in StarCraft II \cite{samvelyan19smac}. In existing value-based MARL algorithms, to adapt to the constraint of agents' partial observation during inference stage, they make the same state setting for each agent during training. In this way, for each agent, its individual Q-value is conditioned only on individual local observation and its action. To take the advantage of the global state and team reward, a mixing network is designed to aggregate the individual Q-values of all agents so as to approximate the team reword following the Individual-Global-Max (IGM) principle~\cite{son2019qtran}. Following this paradigm, there are many algorithms with different designs on the mixing network. For instance, Value DecompositionNetwork (VDN), the first attempt in this field, adopts the summation operation~\cite{sunehag2017value}. Differently, QMIX~\cite{rashid2018qmix} assigns the nonnegative weights to individual Q-values with a non-linear function of the global state. Later, more and more algorithms employ the variants of attention mechanism as the aggregation function~\cite{son2019qtran,yang2020qatten,wang2020qplex}.The algorithms discussed above all suffer an essential limitation, \emph{i.e.,} the individual Q-value estimation of all agents fails to consider the global state, which makes them difficult to learn a tacit coordination. This is due to the information gap between the training phase and inference phase. In other words, during training, the global state is available to all agents. But in inference, only local and partial observation is accessible to each agent. 

To bridge the above gap, in this work, we propose a novel framework, named Centralized Teacher with Decentralized Student (CTDS), with a teach model and a student model.  Specifically, the teach model estimate the individual Q-value estimation of agents based on the global state, while the student model regress each agent's Q-value with only local and partial observation. With global state, the teacher model is likely to learn a more tacit coordination among agents, which is transfered to the student model  with a knowledge distillation strategy.  After the training, the student model is ready to be used for online inference with only local observation. It is notable that our CTDS framework is generic and is ready to be applied upon various existing CTDE methods to boost their performance.

We evaluate our CTDS on a grid world environment Combat and a range of unit micromanagement tasks built in StarCraft II.
Our experiments show that CTDS significantly outperforms the recent representative CTDE methods, in terms of both performance and learning speed.
Moreover, we also investigate the performance of CTDS under the scenario with different sight ranges, which further illustrates the robustness of our method.
\begin{figure}[t]
	\centering
	\includegraphics[width=0.8\columnwidth]{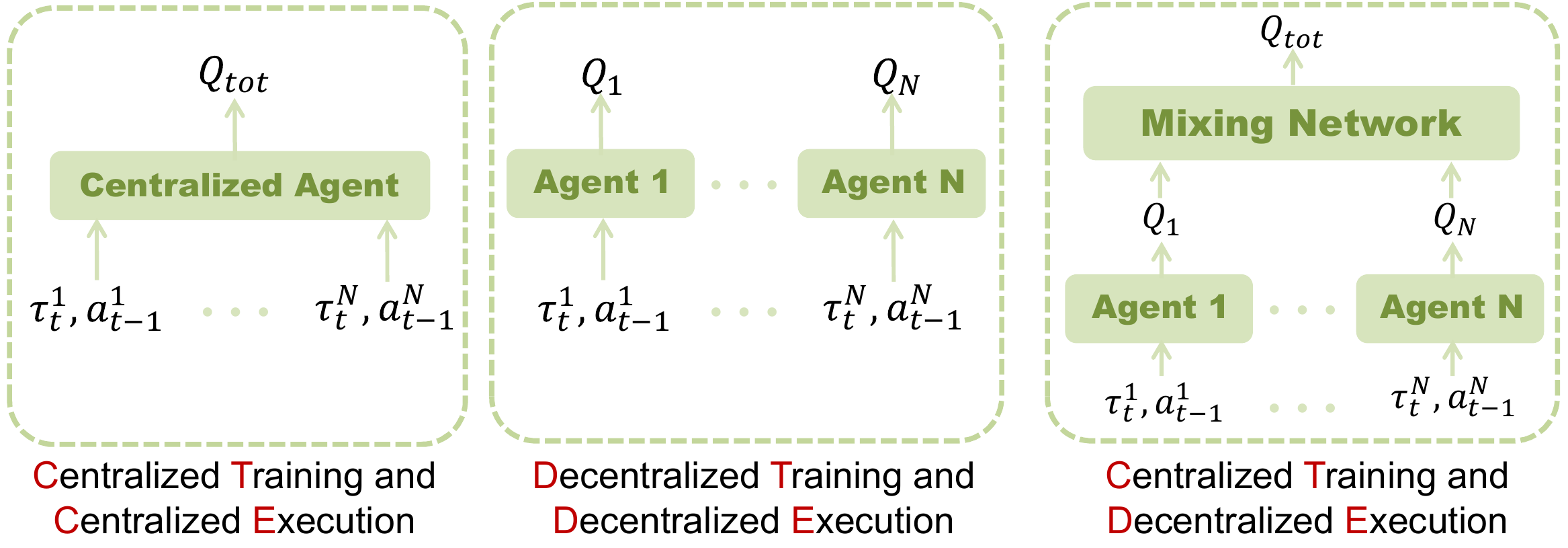}
	\caption{An overview of the Centralized Training and Centralized Execution (CTCE), Decentralized Training and Decentralized Execution (DTDE) and Centralized Training and Decentralized Execution (CTDE) framework.}
	\label{dtde}
\end{figure}

\section{Related Work}
According to the relationship between agents, there are cooperative \cite{oroojlooyjadid2019review} and competitive \cite{t2003decommitment} multi-agent systems.
Due to its wide application in practice, cooperative multi-agent reinforcement learning has gained substantial research interests~\cite{li2021structured}.
In a cooperative multi-agent system, a team of agents coordinate with each other and obtain an overall team reward.
In view of the coordination among agents, early efforts focus on centralized learning and execution~\cite{sukhbaatar2016learning,peng2017multiagent,openai2019dota,ye2020towards}, in which agents can communicate with each other to access full observation of the environment.
However, in many real-world scenarios, the partial observability and/or communication constraints among agents necessitate the decentralized execution.

Considering the necessity of decentralized execution, a paradigm of centralized training with decentralized execution (CTDE) has been widely adopted recently.
To the best of our knowledge, value decomposition network (VDN)~\cite{sunehag2017value,wang2021towards} is the first attempt to allow for centralized value-function learning with decentralized execution.
This work assumes that each agent contributes equally to the team reward, thus decomposes a central action-value function into a sum of individual Q-values conditioned only on local observations.
While achieving decentralization, VDN obtains a sub-optimal solution by simple summation integration and ignores the available global state information during training.

To distinguish the importance of different agents, QATTEN~\cite{yang2020qatten} utilizes a multi-head attention structure to weight the individual Q-values based on the global state and the individual features, and then linearly integrates these values into the central action value.
Unlike VDN and QATTEN that take linear monotonic value functions, QMIX~\cite{rashid2018qmix,rashid2020weighted} relaxes this assumption, and employs a mixing network by leveraging state information to decide the transformation weights, which allows non-linearity operations to decompose the central Q-value.
Later on, QTRAN~\cite{son2019qtran} releases the monotonicity restriction in QMIX\cite{NEURIPS2019_f816dc0a} and proposes a factorization method to express function space induced by IGM consistency.
The computational intractability, however, renders the poor performance of QTRAN in complex tasks.
QPLEX~\cite{wang2020qplex} decomposes the central Q-value into the sum of individual value functions and a non-positive advantage function, which is 0 only when all of the agents choose the optimal actions.
By introducing the duplex dueling structure, QPLEX achieves the complete function class that satisfies the IGM principle.

As far as we know, all the existing CTDE methods only utilize local observations instead of the global observation for learning individual Q-values during training in order to achieve decentralized execution.
Intuitively, in a cooperative multi-agent system, it is beneficial when knowing more about other agents.
Inspired by the benefit of the global observation, we adopt the concept of knowledge distillation~\cite{garcia2020learning,wang2021adversarial}. 

Knowledge distillation is firstly proposed for network compression~\cite{hinton2015distilling}, which distills the knowledge ($i.e.$, the output distribution) from a teacher model that is typically large into a smaller student model so that the student model can achieve similar performance as the teacher model. 
Gradually, the concept of knowledge distillation has been generalized into a framework consisting of a teacher model and a student model, where the student aims to imitate the teacher under its guidance.
In other words, knowledge distillation suggests training by matching the student’s predictions to the teacher’s predictions~\cite{kim2016sequence,phuong2019towards}.
Due to the advantage of fast optimization, network minimization and comparable performance~\cite{yim2017gift,phuong2019towards}, the technique of knowledge distillation has been widely applied in various ways~\cite{kim2016sequence,chen2017learning,mirzadeh2020improved}.
In~\cite{rusu2016policy}, knowledge distillation is firstly leveraged to reinforcement learning to distill the policies to a dramatically smaller and efficient network.
The experiments show that the distilled agents outperform their teachers in most games, which indicates the effectiveness of knowledge distillation in reinforcement learning tasks. 

By leveraging knowledge distillation to MARL, we aims to take full advantage of full observation during training with a teacher module.
To allow decentralized execution, we design the student module, which approximates individual Q-values based on local observations, to distill the Q-values estimated by the teacher module.
In this way, a tacit coordination can be learned among the agents at the student side. 

\begin{figure}[t]
	\centering
	\subfigure[CTDE Framework.]{
		\includegraphics[width=0.33\textwidth]{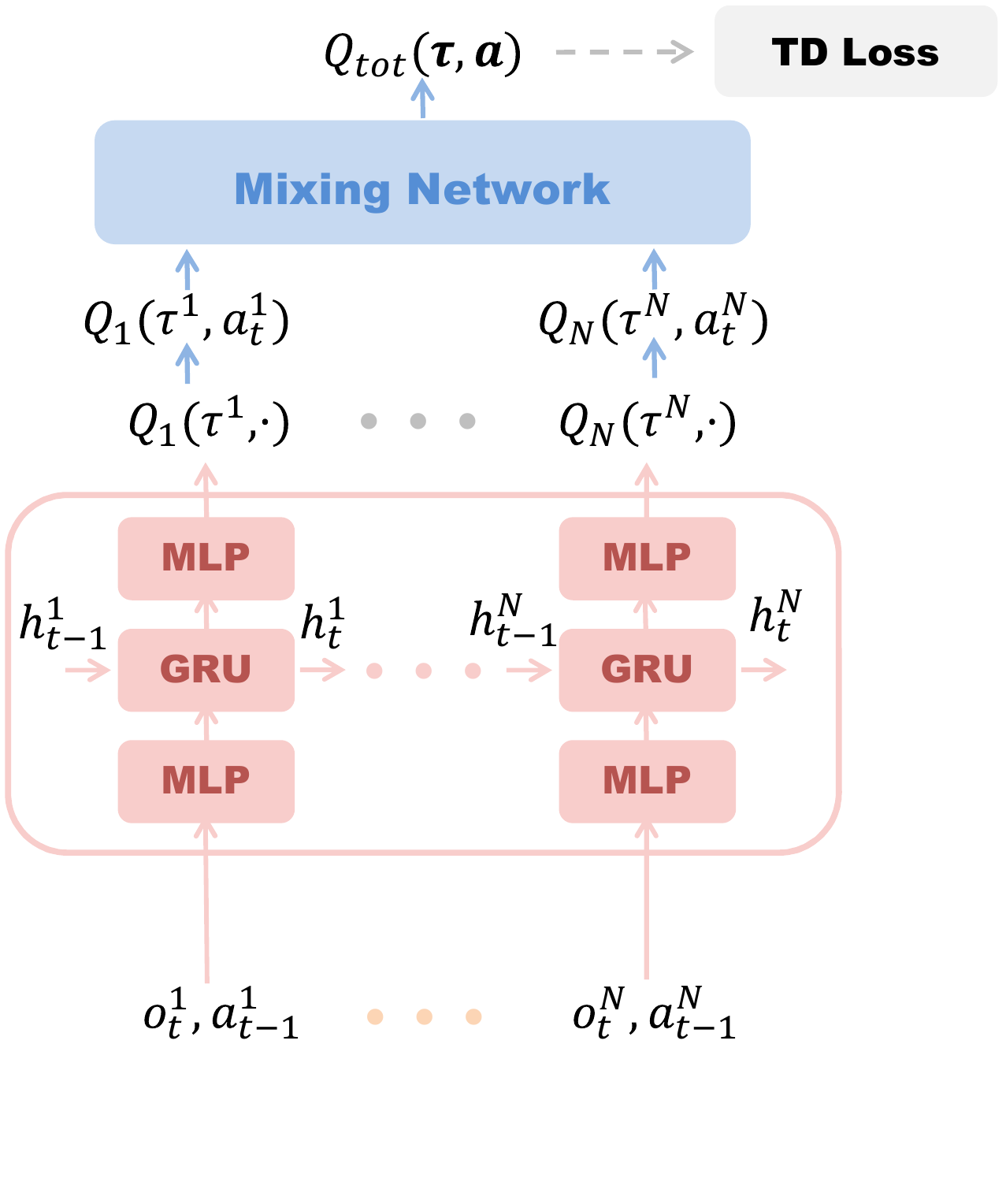}
	}
	\subfigure[CTDS Framework.]{
		\includegraphics[width=0.64\textwidth]{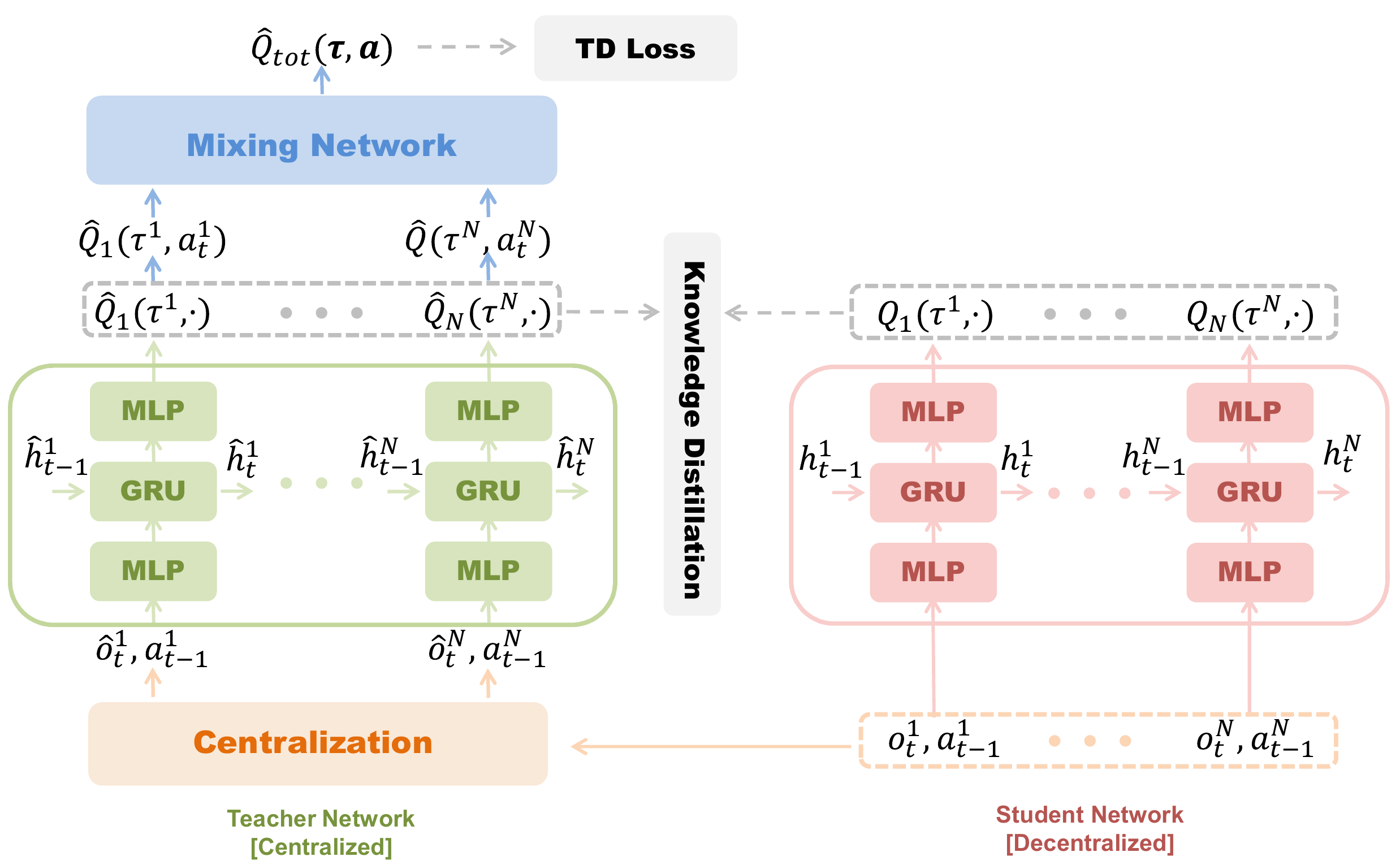}
	}
	\caption{An overview of the CTDE framework and CTDS framework. CTDS consists of the teacher module (on the left) and the student module (on the right).}
	\label{CTDS}
\end{figure}

\section{Preliminaries}
\subsection{Decentralized Partially Observable MDP}
A cooperative multi-agent problem under the partial observability can be modeled as DEC-POMDP~\cite{oliehoek2008optimal,oliehoek2016concise}, which
is defined as a tuple $M = < \bm{N}, \bm{S}, \bm{A}, P, \bm{\Omega}, O, r, \gamma > $, where $\bm{N} \equiv \{ g_1, g_2, \cdots, g_n\} $ is a finite set of agents and $\bm{S}$ describes a set of the global state $s$ of the environment.
At each time step, every agent $g_i \in \bm{N}$ receives an individual partial observation $o_i \in \bm{\Omega}$ according to the observation probability function $O_i(o_i|s)$.
Each agent $g_i$ has an action-observation history $\tau_i \in \bm{\tau} \equiv (\bm{\Omega} \times \bm{A})^*$ and its individual policy $\pi_i(a_i|\tau_i)$.
According to $\pi_i$, each agent $g_i$ chooses an action $a_i \in \bm{A}$, which forms a joint action $\bm{a} \equiv [a_i]_{i=1}^n$.
After the action execution of all agents, it results in a team reward $r(s, \bm{a})$ and a transition to the next global state $s' \sim P(\cdot|s, \bm{a}) $ by the environment.
$\gamma \in [0,1)$ is a discount factor.
The formal objective is to find a joint policy $ \bm{\pi} = <\pi_1,\cdots, \pi_n>$ that maximizes a
joint state value function $V^{\pi}(s) = \mathbb{E} [ \sum_{t=0}^{\infty} \gamma^t r_{t}|s_0 = s, \bm{\pi}]$.
In addition, the joint action-value function $Q^{\bm{\pi}}(s,\bm{a}) = r(s,\bm{a}) + \gamma \mathbb{E}_{s'}[V^{\bm{\pi}}(s')]$ often appears in algorithm expressions to replace the joint state value function $V^{\bm{\pi}}(s)$.

\subsection{Deep Multi-Agent Q-Learning of CTDE Paradigm}
CTDE is a popular paradigm of cooperative multi-agent reinforcement learning~\cite{sunehag2017value,rashid2018qmix,son2019qtran,yang2020qatten,wang2020qplex,oliehoek2008optimal}.
In CTDE, each agent can access the global information during the centralized training, while each agent takes actions only based on local action-observation histories during the decentralized execution. 
Multi-agent Q-learning algorithms represent the individual action-value function of the agent $g_i$ with a neural network parameterized by $\theta_i$, denoted as $Q_i(\tau_i,a_i|\theta_i)$.
Define $\bm{\theta} = [\theta_i]_{i=1}^{N}$ as the parameters of all individual action-value functions.
These algorithms combine the individual action-value $Q_i(\tau_i,a_i|\theta_i)$ with a mixing network $f$ to fit the joint action-value $Q_{tot}$, \emph{i.e.},
\begin{equation}
    Q_{tot} = f(Q_1, Q_2, \cdots, Q_n, s |\phi), 
\end{equation}
where $\phi$ is the parameters of mixing network $f$.
An important concept of CTDE framework is that the mixing network should satisfy the requirement that the optimal joint action induced from the optimal centralized action-value function is equivalent to the collection of individual optimal actions of agents, which is called Individual-Global-Max (IGM), such that the following holds:
\begin{equation}
\arg \max_{\bm{a}} Q_{tot}(\bm{\tau},\bm{a}) = 
\begin{bmatrix} \arg \max_{a_1}Q_1(\tau_1,a_1) \\
\vdots\\
\arg \max_{a_n}Q_n(\tau_n,a_n)
\end{bmatrix}.
\end{equation}

Given the value decomposition, multi-agent Q-learning algorithms use a replay memory $D$ to store the transition tuple $(\bm{\tau}, \bm{a}, r, \bm{\tau'})$, where $r$ is the team reward for taking action $\bm{a}$ at joint action-observation history $\bm{\tau}$ with a transition to $\bm{\tau'}$.
Based on the above discussion, parameters $\bm{\theta}$ and $\phi$ are learnt by minimizing the following expected TD error:
\begin{equation}
\begin{split}
\label{eq:TD}
L(\bm{\theta},\phi) = 
\mathbb{E}_{(\bm{\tau}, \bm{a}, r, \bm{\tau'}) \in D} [(r + \gamma V(\bm{\tau'}|\bm{\theta}^-, \phi^-) -Q_{tot}(\bm{\tau}, \bm{a}|\bm{\theta}, \phi))^2],
\end{split}
\end{equation}
where $V(\bm{\tau'}|\bm{\theta}^-, \phi^-) = \max_{\bm{a'}}Q_{tot}(\bm{\tau'},\bm{a'}|\bm{\theta}^-,\phi^-)$ is the one-step expected future return of the TD target.
$\bm{\theta}^-$ and $\phi^-$ are the parameters of the target network, which will be periodically updated with $\bm{\theta}$ and $\phi$.

\section{Method}
In this section, we first explain the motivation of our proposed framework and introduce the main components in detail.
Besides, we give a theoretical analysis about why our method works.
\subsection{The CTDS Framework}
In this subsection, we present a new framework called Centralized Teacher with Decentralized Student (CTDS), which leverages the idea of knowledge distillation to balance the advantage of global observation and the requirement of decentralized execution.

Recall that in a cooperative multi-agent system, a group of agents coordinate with each other to take joint actions based on the global state information and will obtain a team reward correspondingly.
To achieve decentralize execution, existing CTDE framework requires the individual Q-values in the lower layer only conditioned on the local action-observation histories, and incorporates the state information into mixing network to integrate the individual Q-values into the overall Q-value function.
Obviously, if an agent could have full observation of other agents, it would be more likely to learn a better policy under the cooperative scenario.
Considering the benefit of centralized observations during training, our CTDS decouples the centralized training and decentralized execution with the combination of the teacher module and the student module, as illustrated in Figure~\ref{CTDS}.
The teacher module is dedicated to allocating the team reward through centralized training, while the student module utilizes the partial observations to approximate the individual Q-value estimated by the teacher model.
In the following, we will give a detailed description about the design of the teacher module and student module, and the knowledge distillation mechanism between them, respectively.

\subsubsection{Teacher Module}
The teacher module consists of two parts: one is the Q-network for estimating individual action-value function for each agent; another is the mixing network for integrating the individual Q-values to be the centralized Q-value.
Note that the teacher module only participates in the centralized training.

The teacher module allows that each agent has an infinite sight range to receive the observation of all the agents.
For each agent $g_i$, the Q-network takes the centralized observation $\hat{o}_i^t$, last action $a_i^{t-1}$ and last hidden state $\hat{h}^{t-1}_i$ as inputs, fits the perfect individual action-value $\hat{Q}_i(\hat{\tau}_i,\cdot)$ through the multi-layer perceptron (MLP) and gated recurrent unit (GRU) modules, and outputs the optimal action-value $\hat{Q}_i(\hat{\tau}_i,a_i^t)$.
Given the perfect individual action-value $\hat{Q}_i(\hat{\tau}_i,a_i^t)$, the mixing network $f$ combines these individual Q-values into the joint action-value function $\hat{Q}_{tot}$ by utilizing the global state information.
The parameters of the teacher module are updated by the TD loss defined in equation~\ref{eq:TD} iteratively.

\subsubsection{Student Module}
To achieve decentralized execution, the agent $g_i$ in the student module only has access to partial observation $o_i^t$, different from the full observation $\hat{o}_i^t$ in the teacher module.
Compared with the teacher module, the student module only contains the Q-network, which has the same network architecture as the one in the teacher module but the parameters of Q-networks are not shared.
Similarly, the Q-network takes the partial observation $o_i^t$, last action $a_i^{t-1}$ and last hidden state $h^{t-1}_i$ as inputs, and generates the individual action-value $Q_i(\tau_i,\cdot)$ through the MLP and GRU modules.
It can be seen that the student module does not involve any global information and has a relatively simple network structure.
\subsubsection{Knowledge Distillation Mechanism}
As introduced above, the teacher module learns to allocate the team reward reasonably by taking full advantage of the global information while the student module focuses on learning a reasonable individual local Q-value to realize decentralized execution.
To bridge the teacher module and the student module, we employ the knowledge distillation mechanism, $i.e.$, distilling the perfect action-value $\hat{Q}_i$ estimated by the teacher model to guide the student module to approximate the local Q-value.
Specifically, we adopt a mean-squared-error loss (MSE) to minimize the difference between the Q-values estimated by these two modules.
The advantage of MSE is that it constrains the complete set of action-value in the student model~\cite{rusu2016policy}.
Formally, the MSE loss is calculated as follows:
\begin{equation}
    L_{\text{MSE}} = \Sigma_{i=1}^{n}\Sigma_{a \in A}(\hat{Q}_i(\hat{\tau_i},a)-Q_i(\tau_i,a))^2.
\end{equation}

\subsubsection{Training and Execution}
In the training phase, the teacher module generates the corresponding actions of each agent based on the centralized observations to interact with the environment.
Formally, the interactive actions are the results of the $\epsilon$-greedy from the $\hat{Q}_i$, which means choosing random actions with probability $\epsilon$, otherwise choosing actions with the maximal $\hat{Q}_i$.
The parameters of the teacher module and the student module are updated at the same time through iterative interactions.
The teacher module is optimized under the constraint of TD loss while the student module learns the parameters with MSE loss.
The execution phase only relies on the decentralized student module which utilizes the local action-observation records. 
To be concrete, each agent chooses a greedy action $a_i^t$ at each timestamp with respect to the individual Q-value $Q_i$ estimated by the student module.
Therefore, the CTDS framework still meets centralized training and decentralized execution.
For better understanding, we provide detailed procedure of our framework below.

\subsection{Theoretical Analysis}\label{sec:anal}
In this subsection, we theoretically derive why the student module performs better than the execution module of CTDE with the equivalent input information. 

As mentioned above, the teacher module learns from the centralized history trajectory $\hat{\tau}_i$, which contains the decentralized information ${\tau}_i$ (accessible to the student module) and the complement information $\tau^*_i=\hat{\tau}_i^t\setminus{\tau}_i^t$ (we assume that $\tau_i$ and $\tau^*_i$ are independent).
In order to explain the relationship between input trajectories and output individual Q-values more clearly, we simplify the student module and teacher module into the following formulation:
\begin{equation}
\begin{aligned}
    (\text{Teacher network})\qquad &f_{T}:(\tau_i, \tau^*_i)\rightarrow \hat{Q}_i,\\
    (\text{Student network})\qquad &f_{S}:\tau_i\rightarrow Q_i.
\end{aligned}
\end{equation}
The student module of CTDS and the execution module of CTDE both employ decentralized information merely. 
Due to the limitation of partial trajectory, $i.e.$, the lack of $\tau^*_i$, the execution module of CTDE tends to make non-optimal actions for the current state.
The centralized trajectory releases the restriction above, however, it's usually unavailable during execution. 

CTDS takes both sides into consideration. For a certain decentralized trajectory ${\tau}_i$, the teacher module can give a centralized Q-value $f_T(\tau_i, \tau^*_i)$ with any possible implicit knowledge $\tau^*_i$. 
If the expectation of teacher module's output on implicit information $\mathbb{E}_{\tau^*_i} f_T(\tau_i, \tau^*_i)$ is achieved, it can be used to instruct student module $f_S$ to reduce environment uncertainty caused by the partial trajectory. 
Nevertheless, approximating $\mathbb{E}_{\tau^*_i} f_T(\tau_i, \tau^*_i)$ directly needs a large number of centralized trajectory samples, which is computationally expensive.

The following theorem proves that CTDS framework successfully employs $\mathbb{E}_{\tau^*_i} f_T(\tau_i, \tau^*_i)$ to train student module. 
The comprehensiveness and stability provided by the instruction from the teacher module is the principal cause of the student module's performance promotion.

\textbf{Theorem 1.} For any decentralized trajectory ${\tau}_i$, the student module of CTDS
$f_{S}(\tau_i)$ is instructed to approximate the expectation of a centralized Q-value function  on implicit information $\mathbb{E}_{\tau^*_i} f_T(\tau_i, \tau^*_i)$.

\emph{Proof:}
With a fixed teacher network, the optimization target of student network is
\begin{equation}
    \mathop{\arg\min}_{\theta_S} \ \ \sum\nolimits_{i}\mathbb{E}_{\tau_i, \tau^*_i}
    [f_S(\tau_i)-f_T(\tau_i, \tau^*_i)]^2,
\end{equation}
where $\theta_S$ is the parameters of the student module. It can be proven that:
\begin{equation}
\begin{aligned}
    & \ \ \ \ \frac{\partial}{\partial \theta_S}\mathbb{E}_{\tau^*_i}
    [f_S(\tau_i)-f_T(\tau_i, \tau^*_i)]^2\\ 
    &= \frac{\partial}{\partial \theta_S}\int [f_S(\tau_i)-f_T(\tau_i, \tau^*_i)]^2 p(\tau^*_i) d\tau^*_i \\
    &= \int 2(f_S(\tau_i)-f_T(\tau_i, \tau^*_i)) p(\tau^*_i) \frac{\partial f_S(\tau_i)}{\partial \theta_S} d\tau^*_i \\
    &= 2\frac{\partial f_S(\tau_i)}{\partial \theta_S}[f_S(\tau_i)- \int f_T(\tau_i, \tau^*_i)) p(\tau^*_i) d\tau^*_i]\\
    &= \frac{\partial}{\partial \theta_S}[f_S(\tau_i)- \int f_T(\tau_i, \tau^*_i)) p(\tau^*_i) d\tau^*_i]^2\\
    &= \frac{\partial}{\partial \theta_S}[f_S(\tau_i)- \mathbb{E}_{\tau^*_i} f_T(\tau_i, \tau^*_i)]^2\\
\end{aligned}
\end{equation}
Therefore, 
\begin{equation}
\begin{split}
  \mathop{\arg\min}_{\theta_S}  \sum\nolimits_{i}\mathbb{E}_{\tau_i, \tau^*_i}
    [f_S(\tau_i)-f_T(\tau_i, \tau^*_i)]^2 
    = \mathop{\arg\min}_{\theta_S}  \sum\nolimits_{i}\mathbb{E}_{\tau_i}
    [f_S(\tau_i)- \mathbb{E}_{\tau^*_i} f_T(\tau_i, \tau^*_i)]^2.
\end{split}
\end{equation}

\section{Experiments}
In this section, we conduct comprehensive experiments to demonstrate the effectiveness of our method. 
All of the following experiments are to prove that our approach is robust enough to work on a variety of tasks.
Our codes are available in https://github.com/cathyhxh/CTDS.

\subsection{Settings}\label{sec:set}

In this paper, we test our method on two environments: Combat (a grid world environment introduced by~\cite{magym}) and StarCraft II micromanagement (SMAC)~\cite{samvelyan19smac}.

\begin{figure}[t]
	\centering
	\subfigure[Combat]{
		\includegraphics[width=0.17\columnwidth]{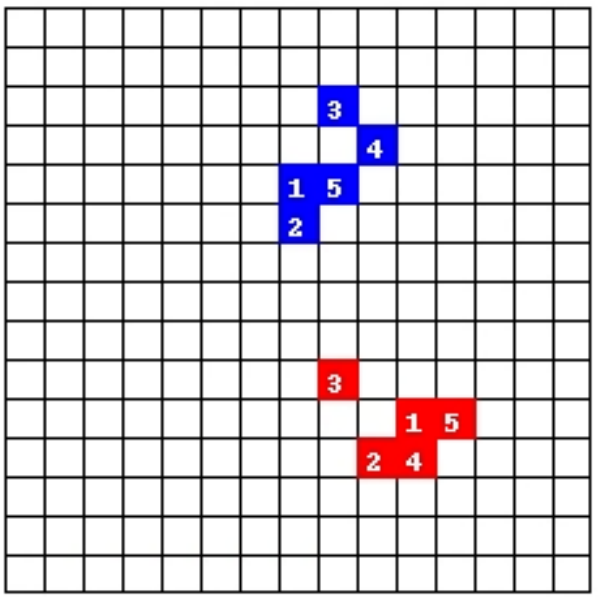}
	}
	\subfigure[SMAC]{
		\includegraphics[width=0.17\columnwidth]{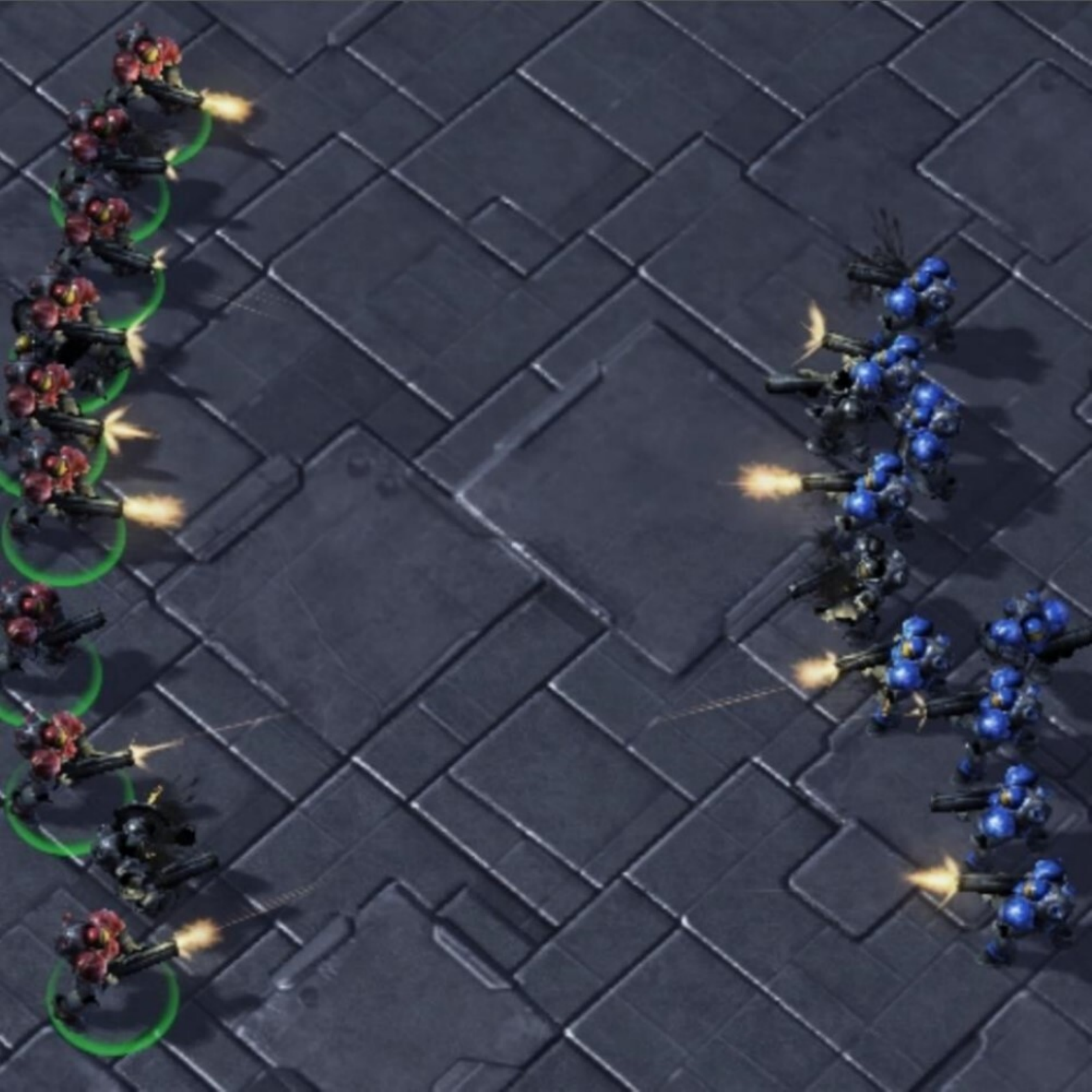}
	}
	\subfigure[with sight range limit]{
		\includegraphics[width=0.25\columnwidth]{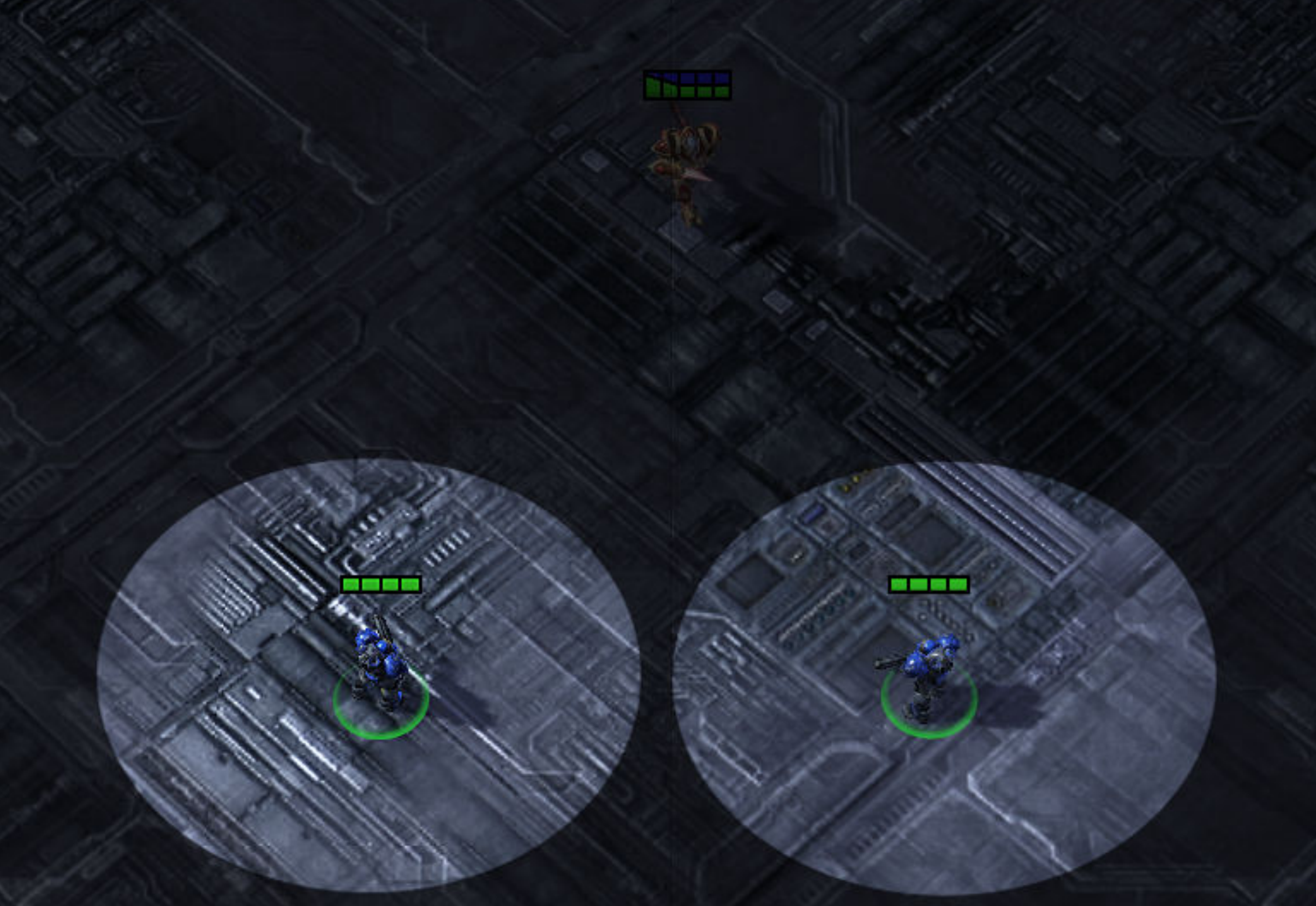}
		\label{fig2-1}
	}
	\subfigure[without sight range limit]{
		\includegraphics[width=0.25\columnwidth]{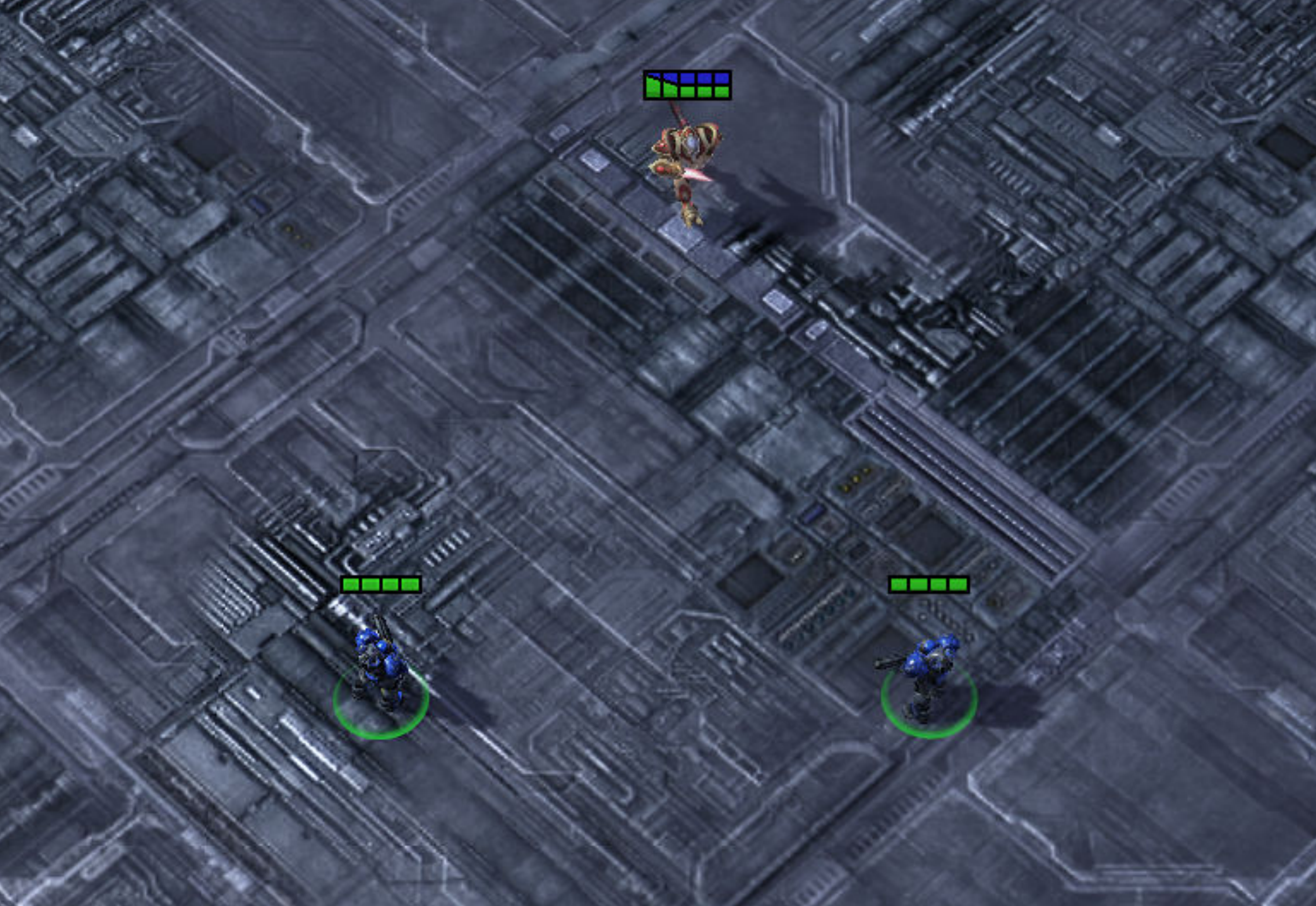}
		\label{fig2-2}
	}
	\label{fig:sight}
	\caption{(a)\&(b) Visualisation of the two experimental environments. (c)\&(d) An illustration about the agents with partial and global observation in SMAC.}
\end{figure}


\begin{figure*}[t]
	\centering
	\subfigure[VDN-5m\_vs\_6m]{
		\includegraphics[width=0.22\textwidth]{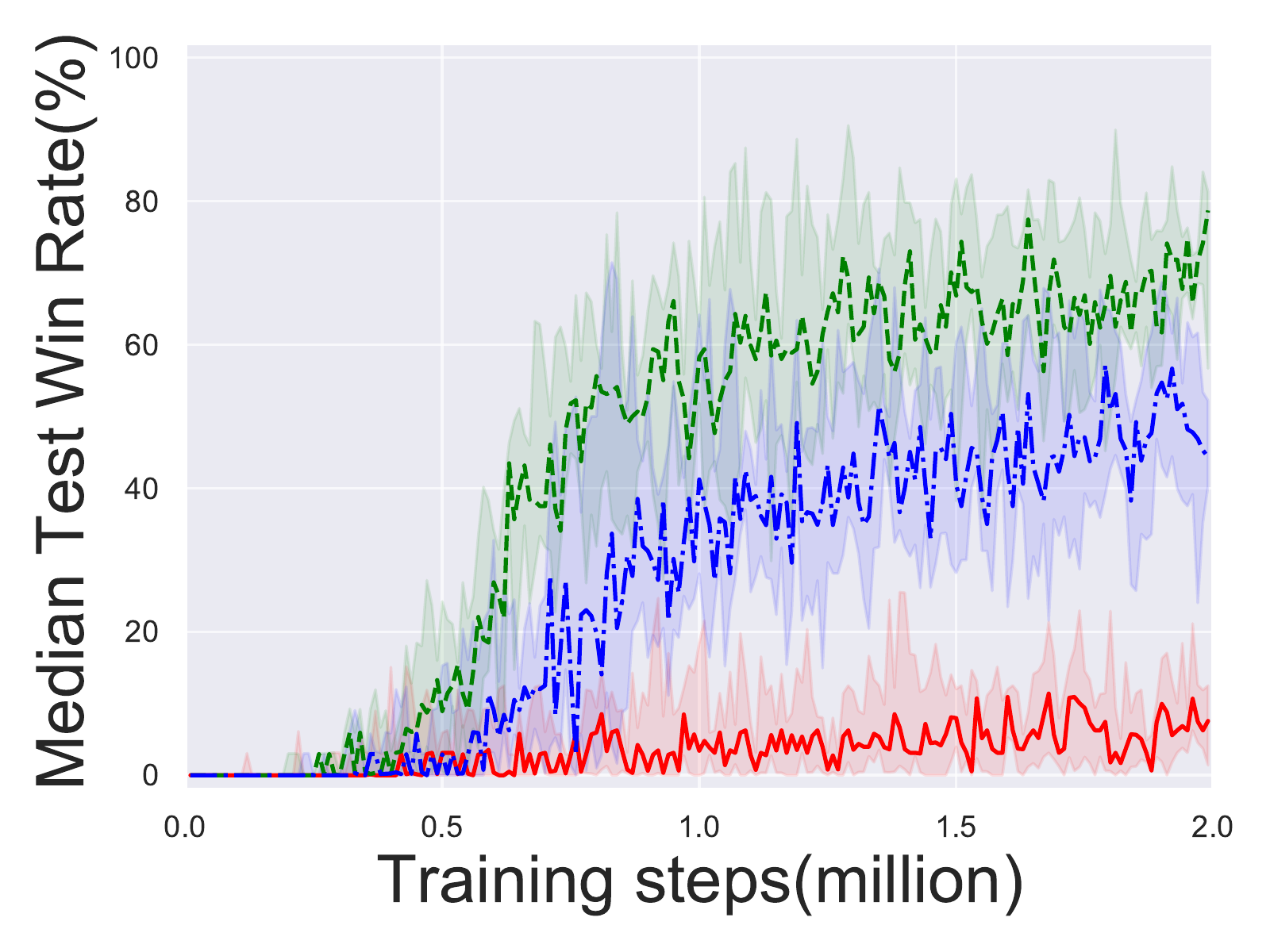}
		\label{VDN-5m_vs_6m}
	}
	\hfill
	\subfigure[QMIX-5m\_vs\_6m]{
		\includegraphics[width=0.22\textwidth]{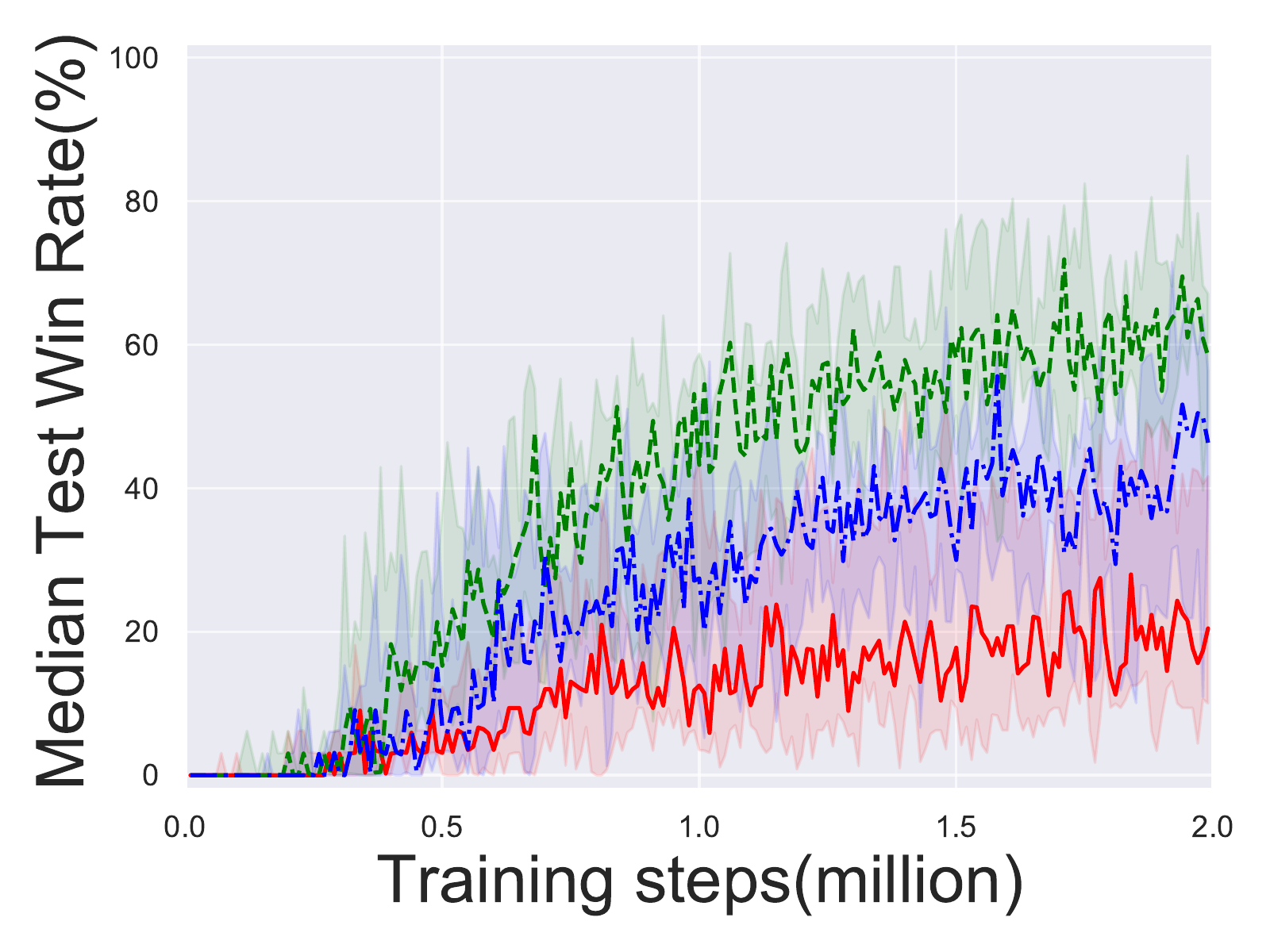}
		\label{QMIX-5m_vs_6m}
	}
	\hfill
	\subfigure[QPLEX-5m\_vs\_6m]{
		\includegraphics[width=0.22\textwidth]{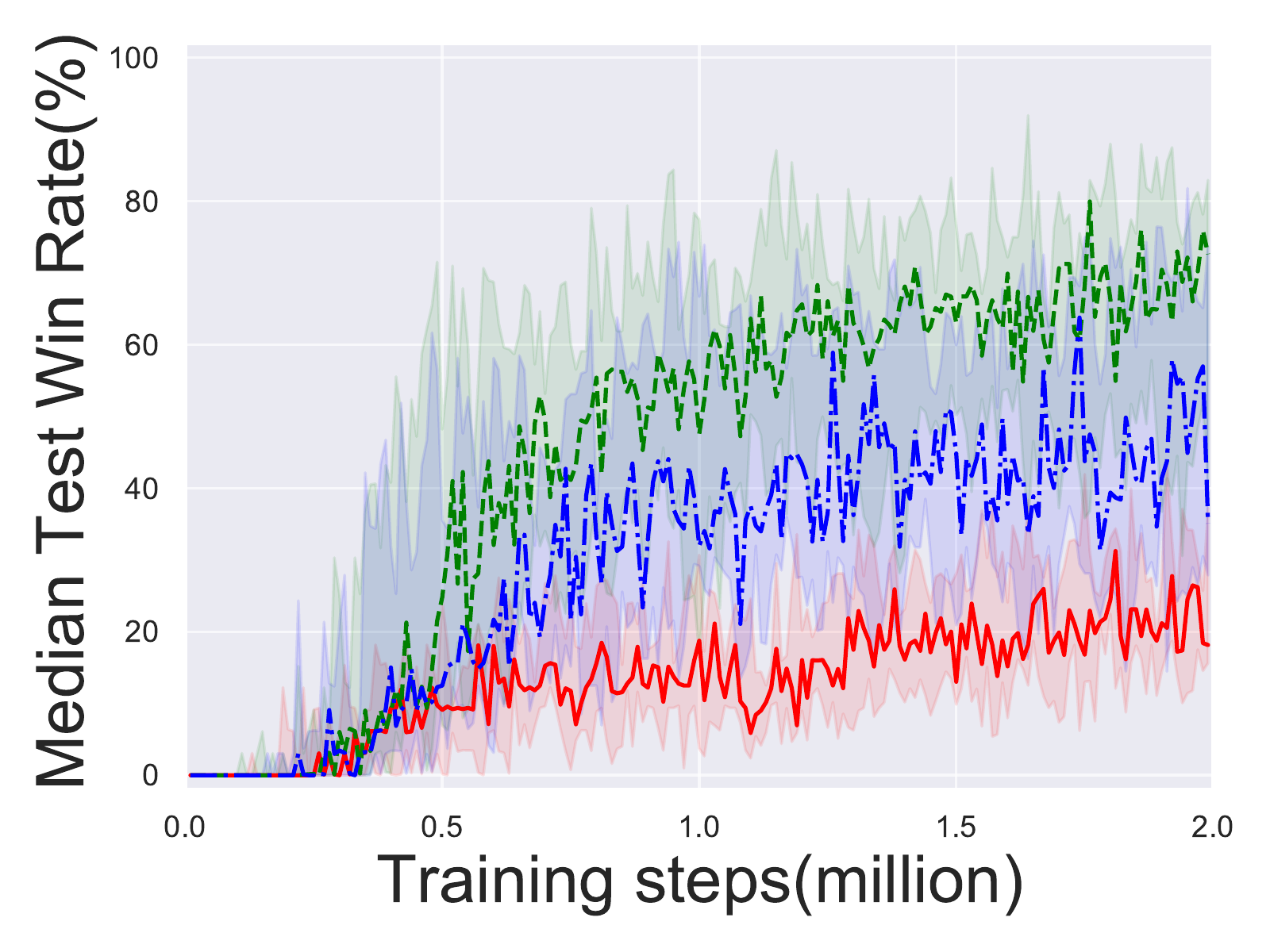}
		\label{QPLEX-5m_vs_6m}
	}
	\hfill
	\subfigure[VDN-bane\_vs\_bane]{
		\includegraphics[width=0.22\textwidth]{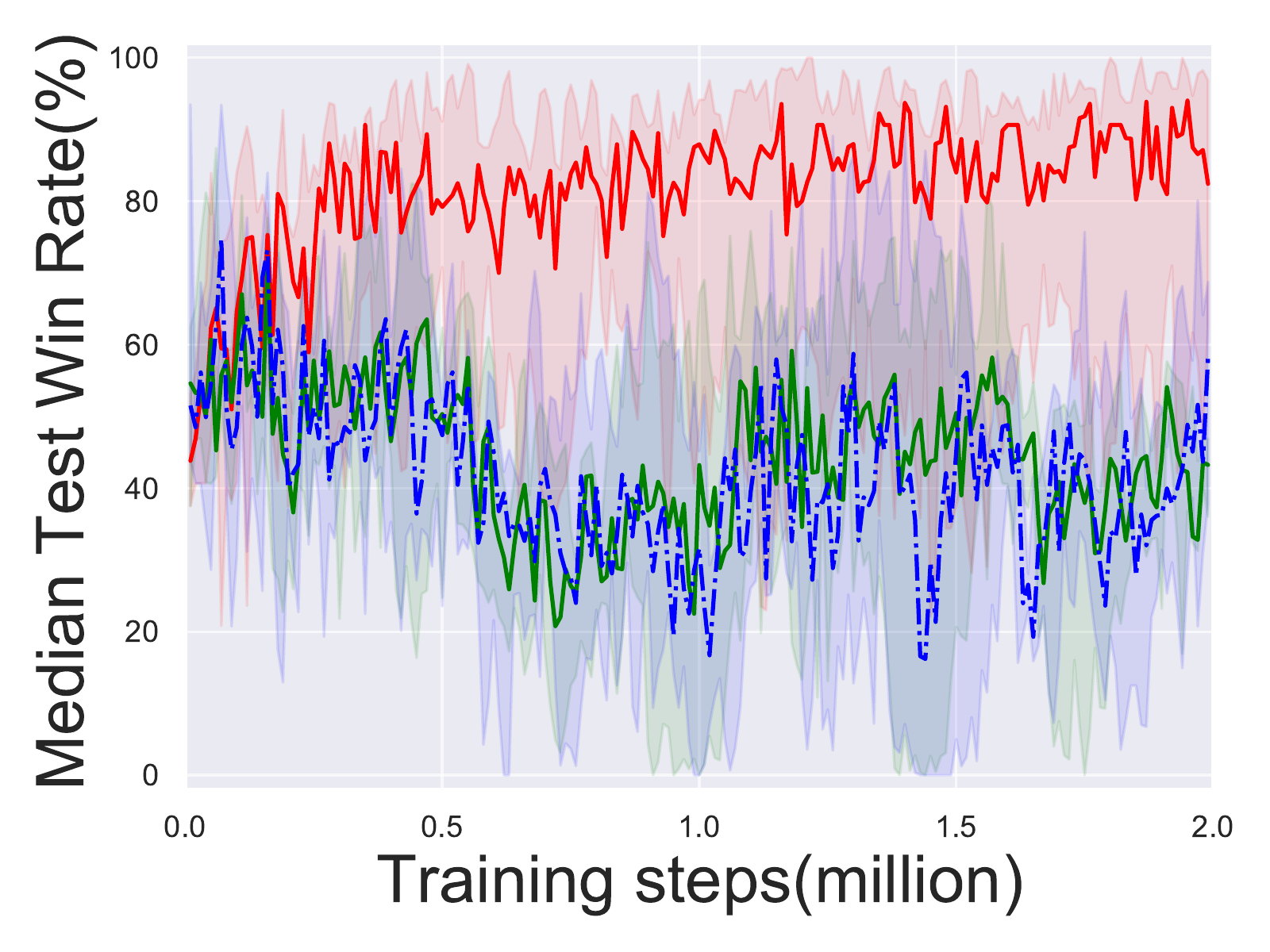}
		\label{VDN-bane_vs_bane}
	}
	\newline
	\subfigure[QMIX-bane\_vs\_bane]{
		\includegraphics[width=0.22\textwidth]{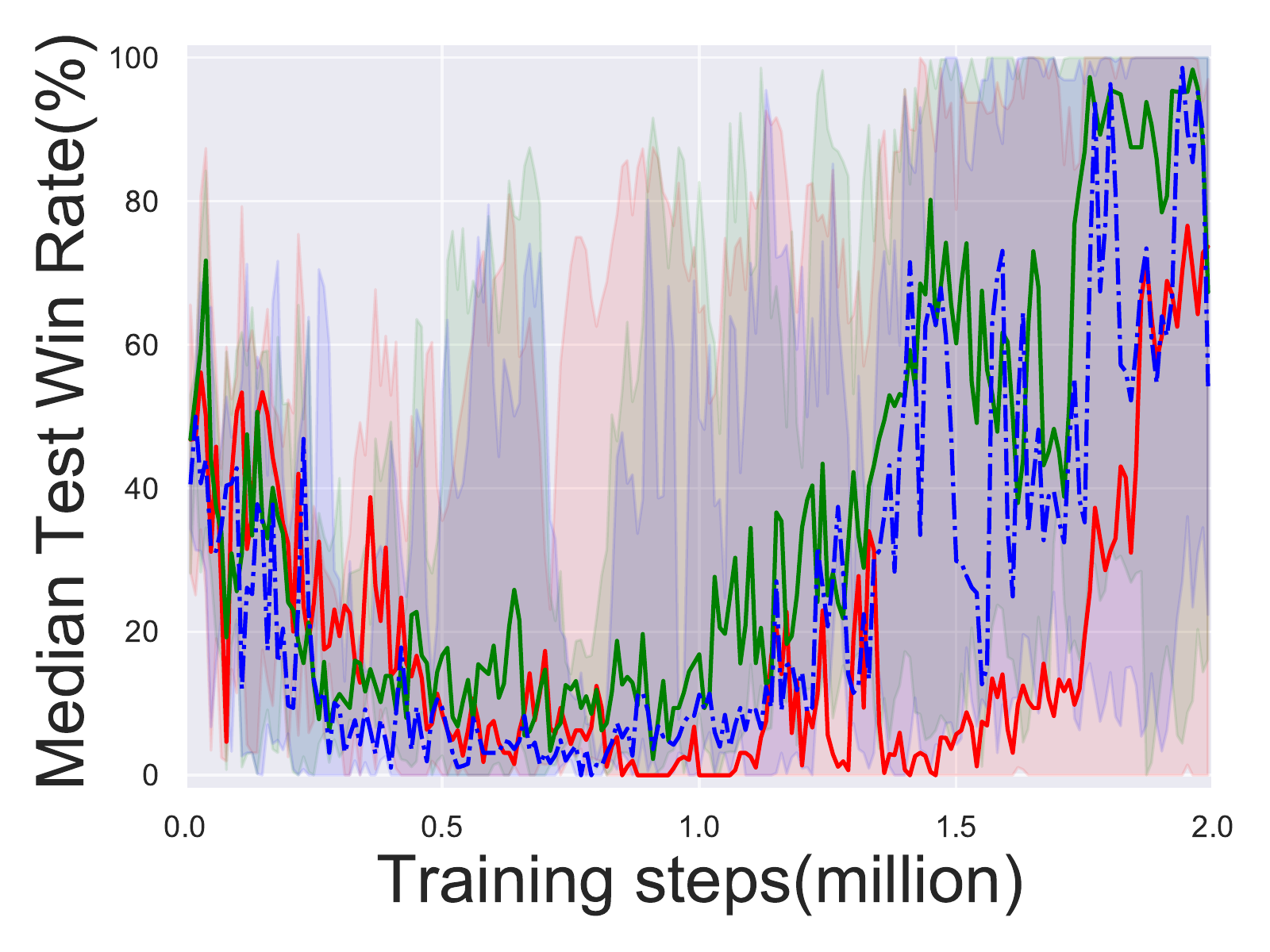}
		\label{QMIX-bane_vs_bane}
	}
	\hfill
	\subfigure[QPLEX-bane\_vs\_bane]{
		\includegraphics[width=0.22\textwidth]{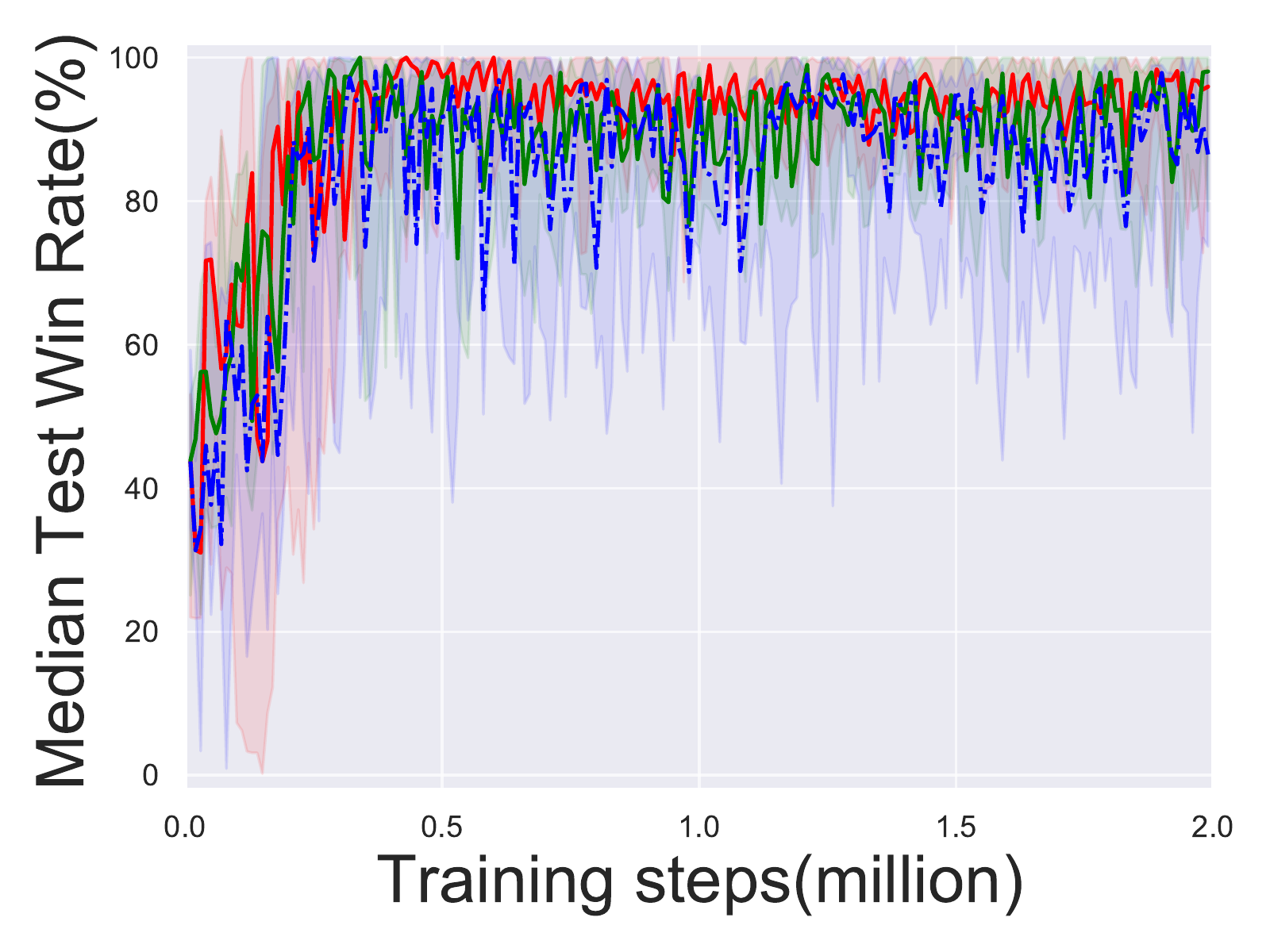}
		\label{QPLEX-bane_vs_bane}
	}
	\hfill
	\subfigure[VDN-27m\_vs\_30m]{
		\includegraphics[width=0.22\textwidth]{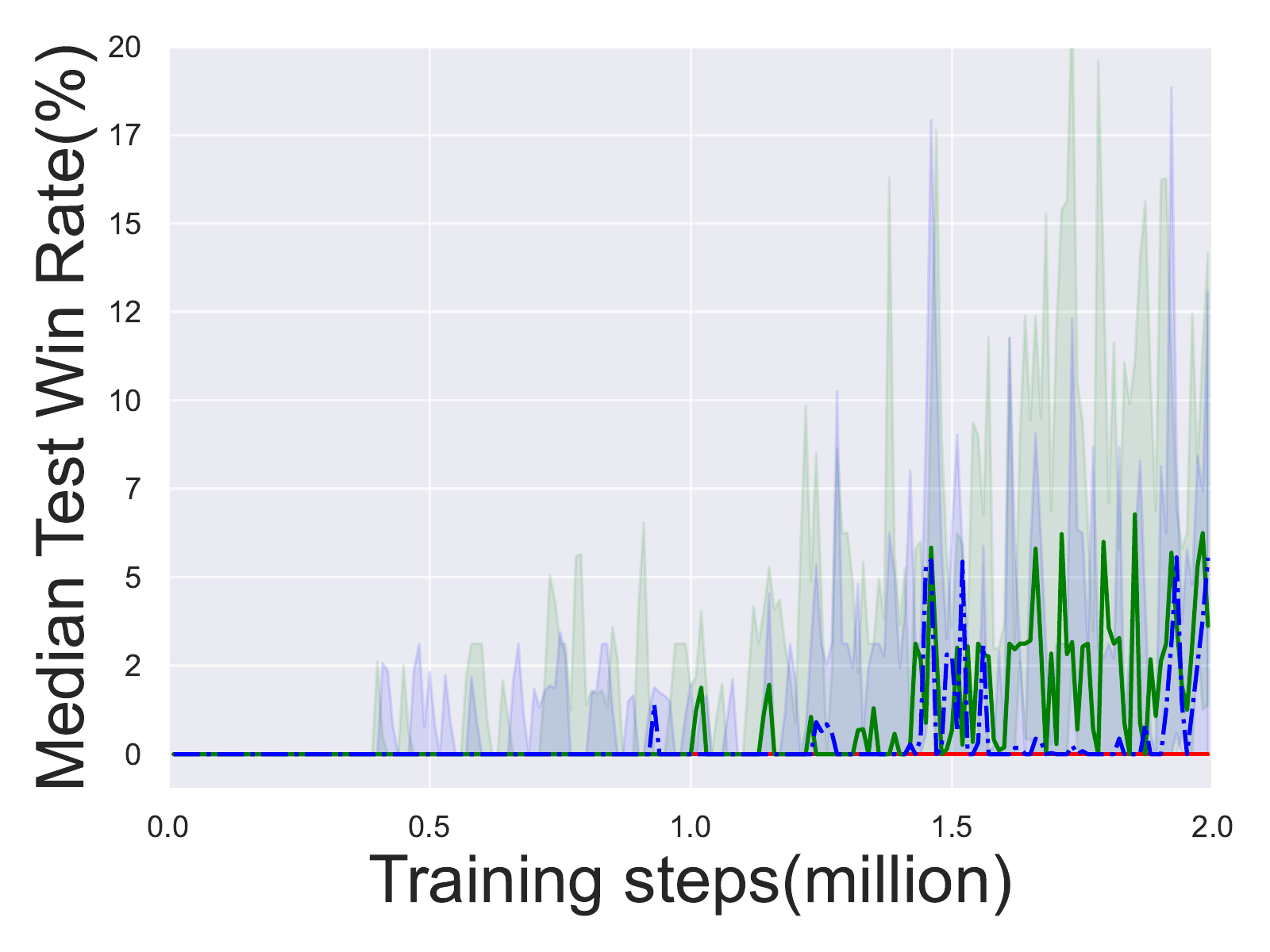}
		\label{VDN-27m_vs_30m}
	}
	\hfill
	\subfigure[QMIX-27m\_vs\_30m]{
		\includegraphics[width=0.22\textwidth]{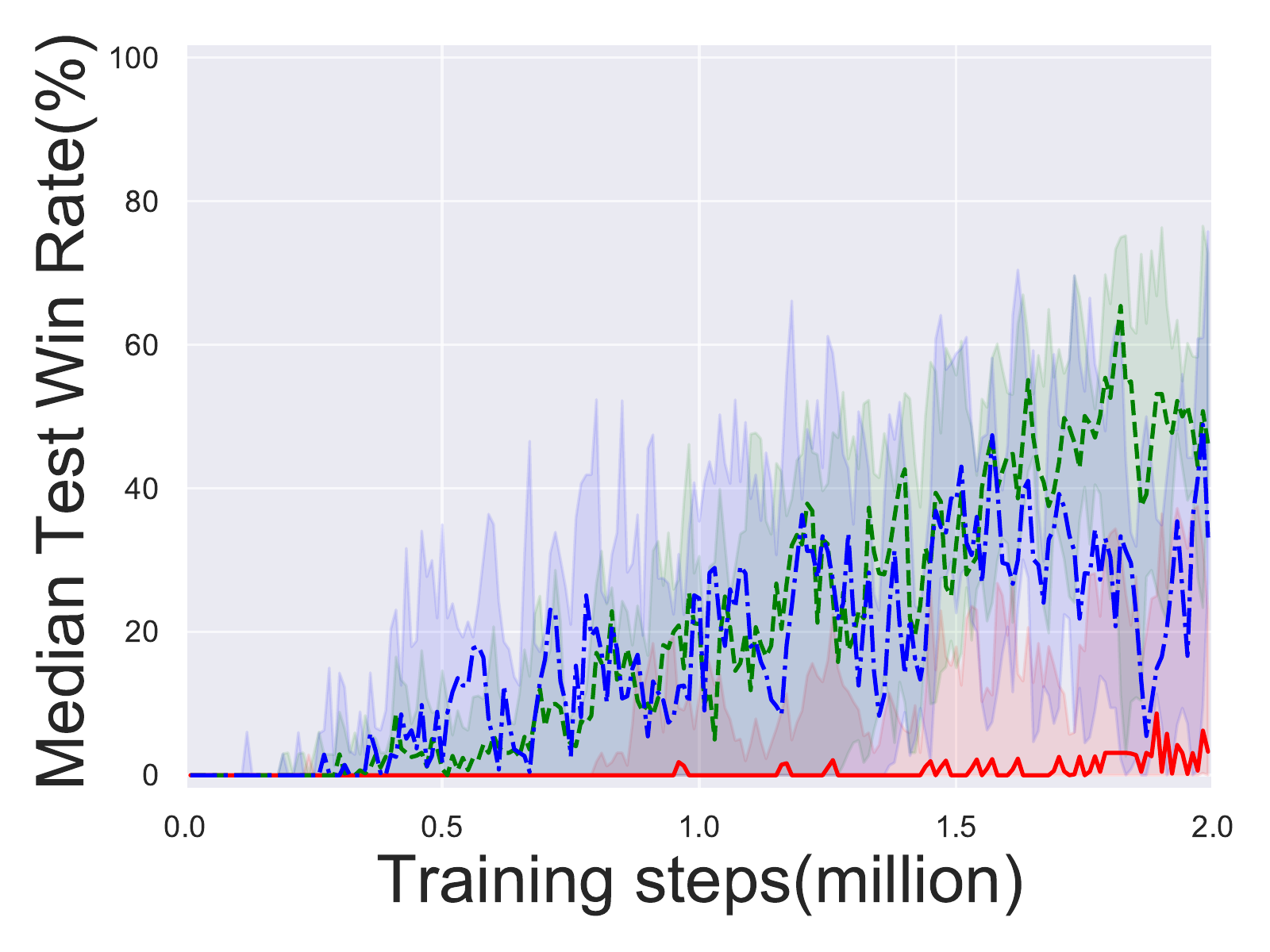}
		\label{QMIX-27m_vs_30m}
	}
	\newline
	\subfigure[QPLEX-27m\_vs\_30m]{
		\includegraphics[width=0.22\textwidth]{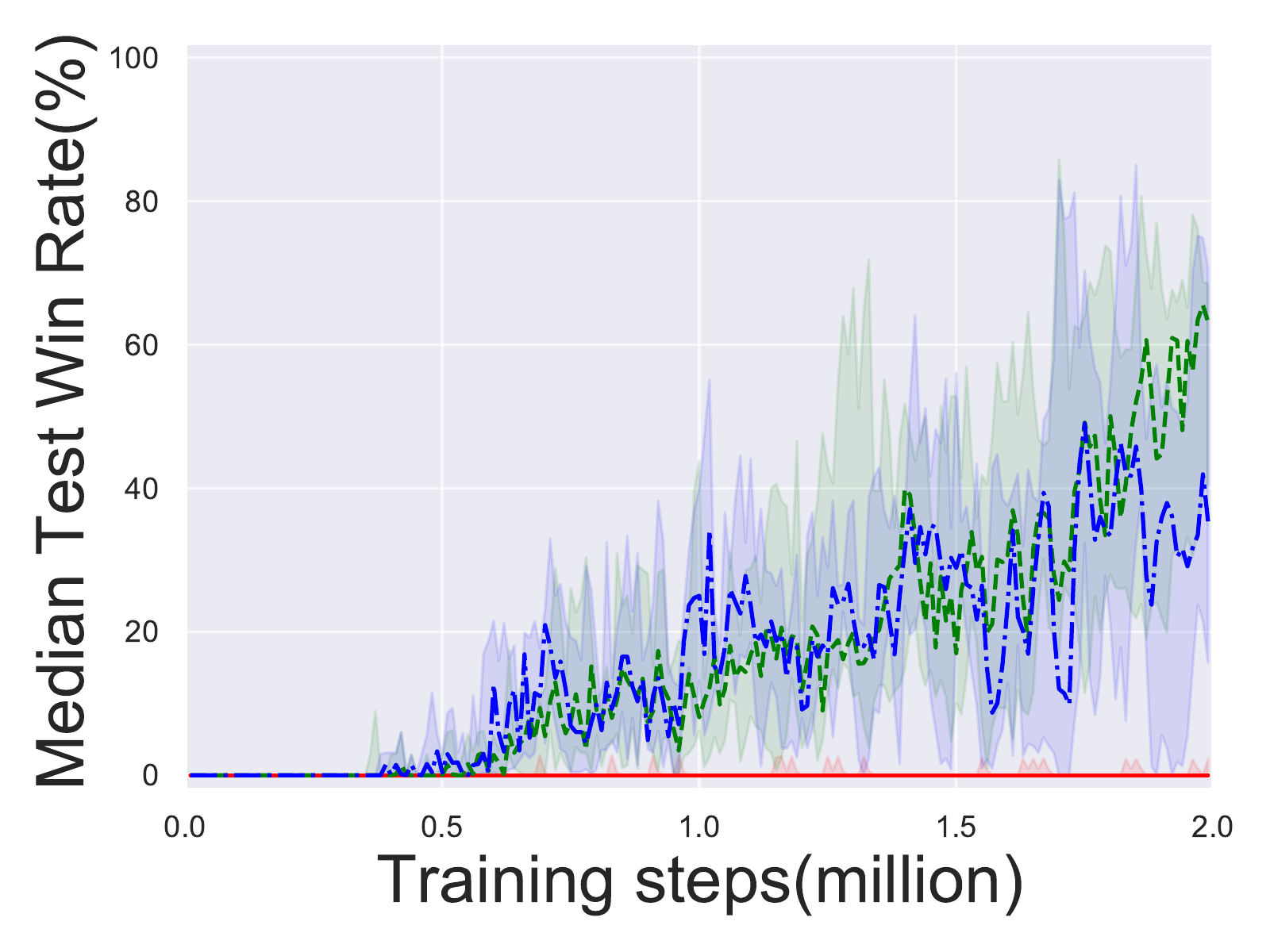}
		\label{QPLEX-27m_vs_30m}
	}
	\hfill
	\subfigure[VDN-MMM2]{
		\includegraphics[width=0.22\textwidth]{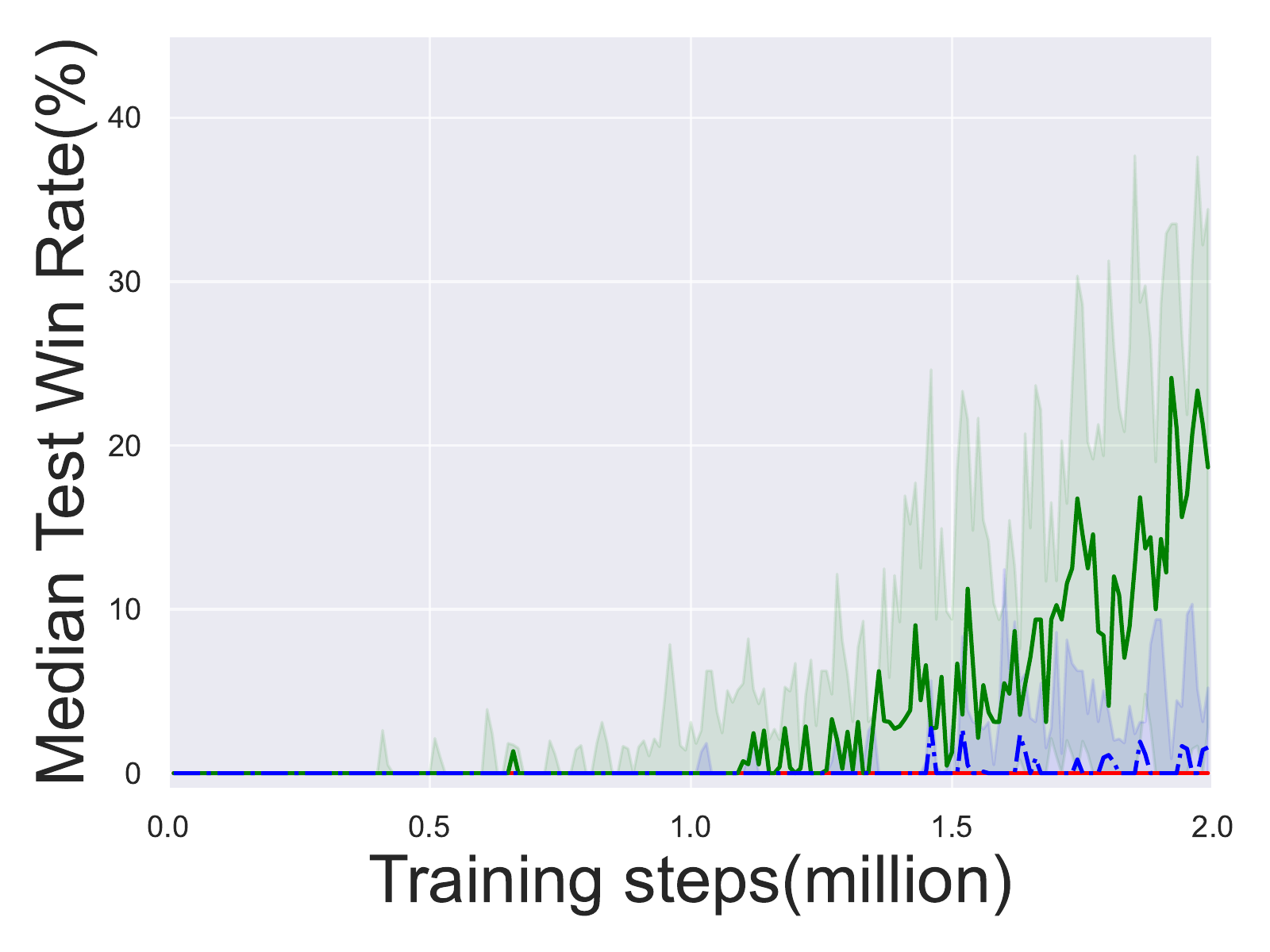}
		\label{VDN-MMM2}
	}
	\hfill
	\subfigure[QMIX-MMM2]{
		\includegraphics[width=0.22\textwidth]{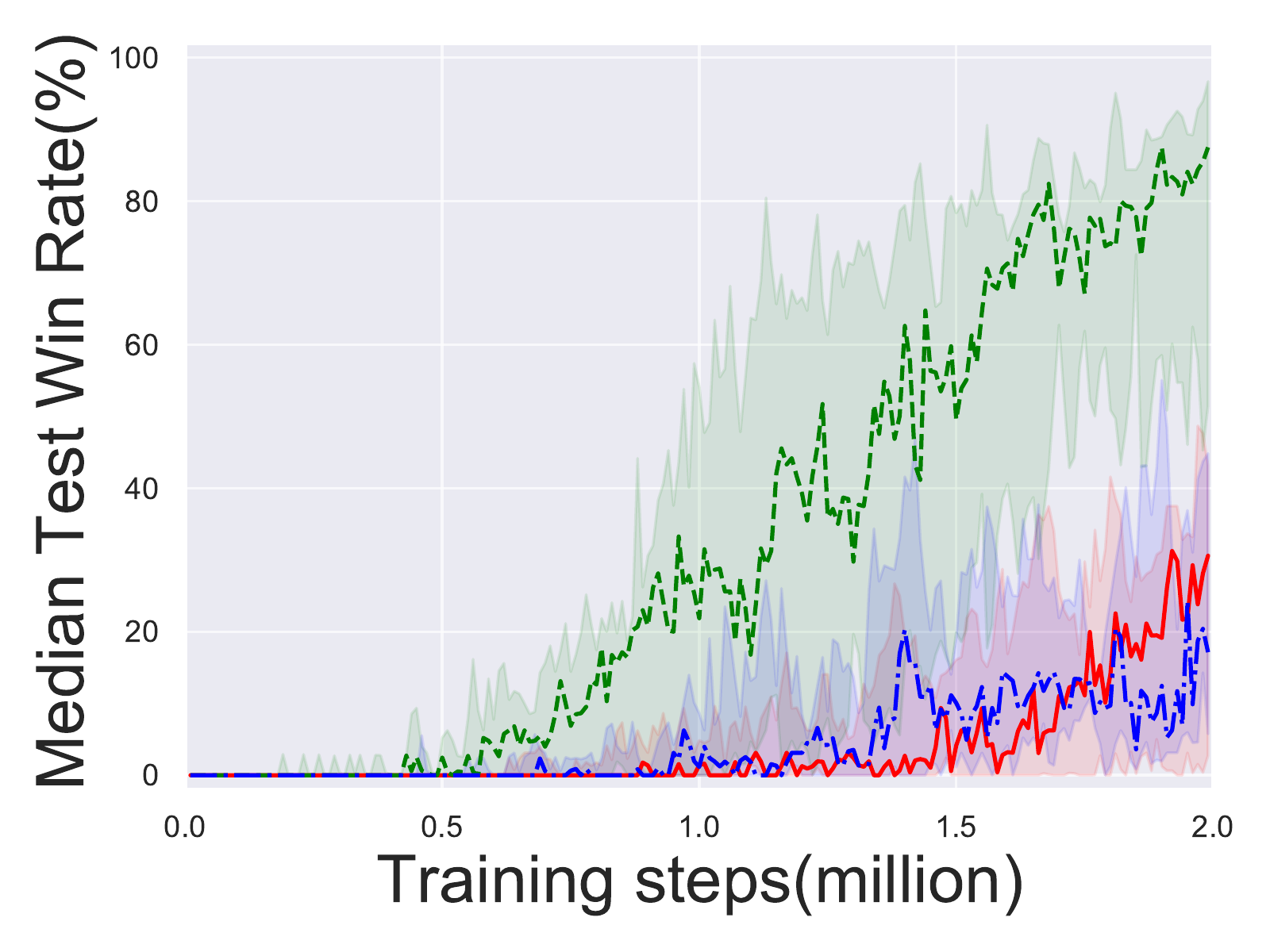}
		\label{QMIX-MMM2}
	}
	\hfill
	\subfigure[QPLEX-MMM2]{
		\includegraphics[width=0.22\textwidth]{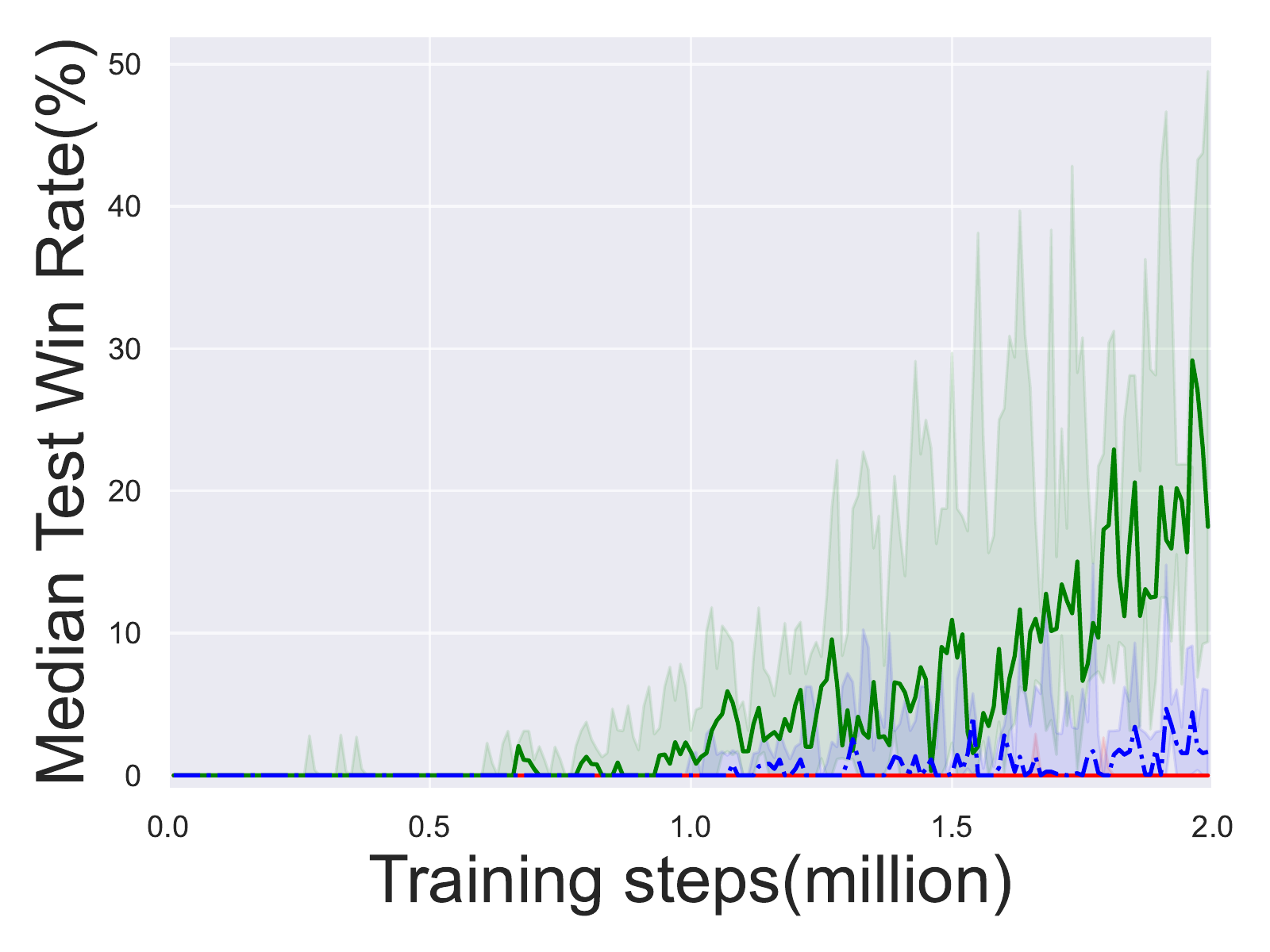}
		\label{QPLEX-MMM2}
	}
	\newline
	\subfigure[VDN-2c\_vs\_64zg]{
		\includegraphics[width=0.22\textwidth]{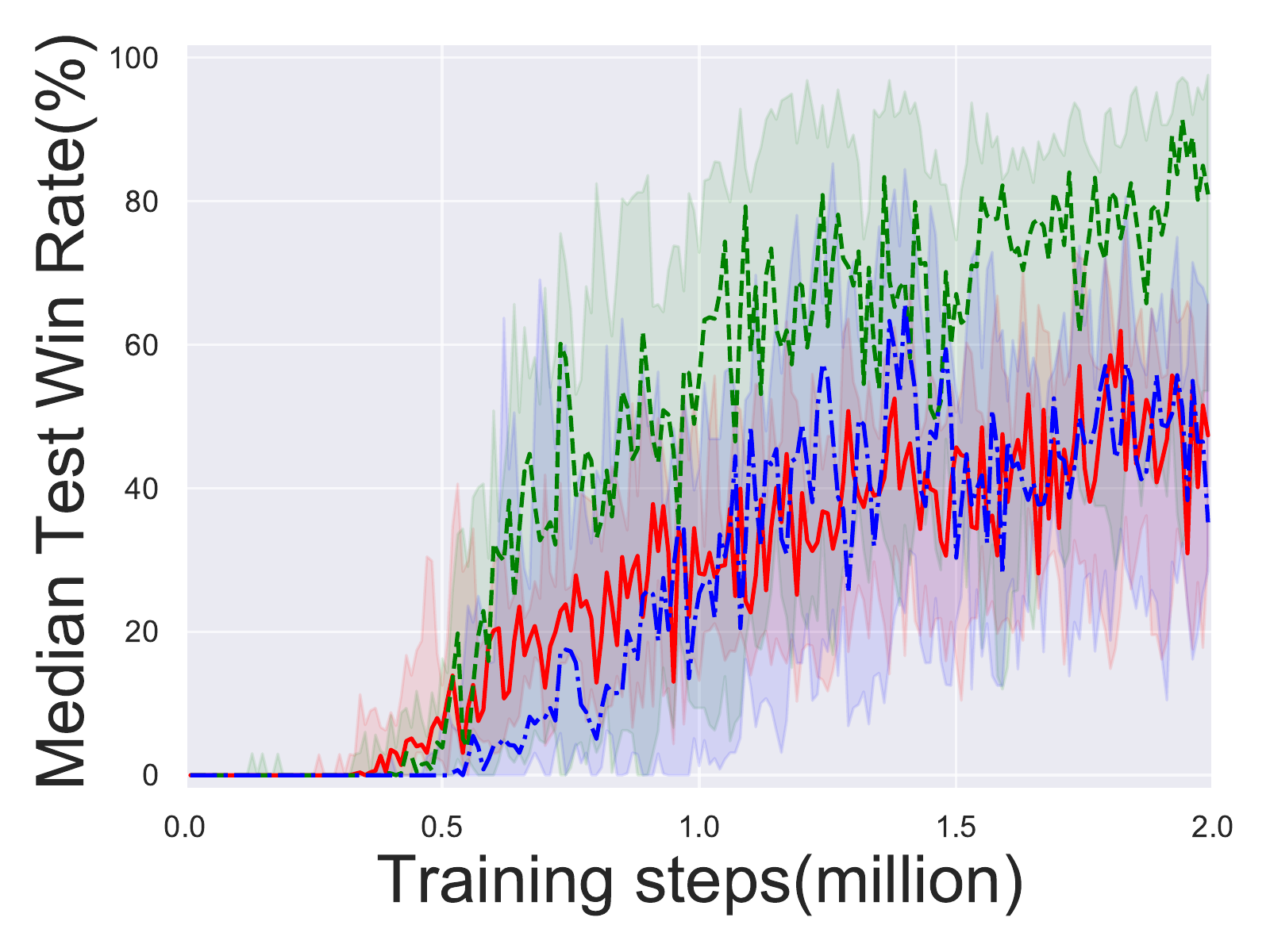}
		\label{VDN-2c_vs_64zg}
	}
	\hfill
	\subfigure[QMIX-2c\_vs\_64zg]{
		\includegraphics[width=0.22\textwidth]{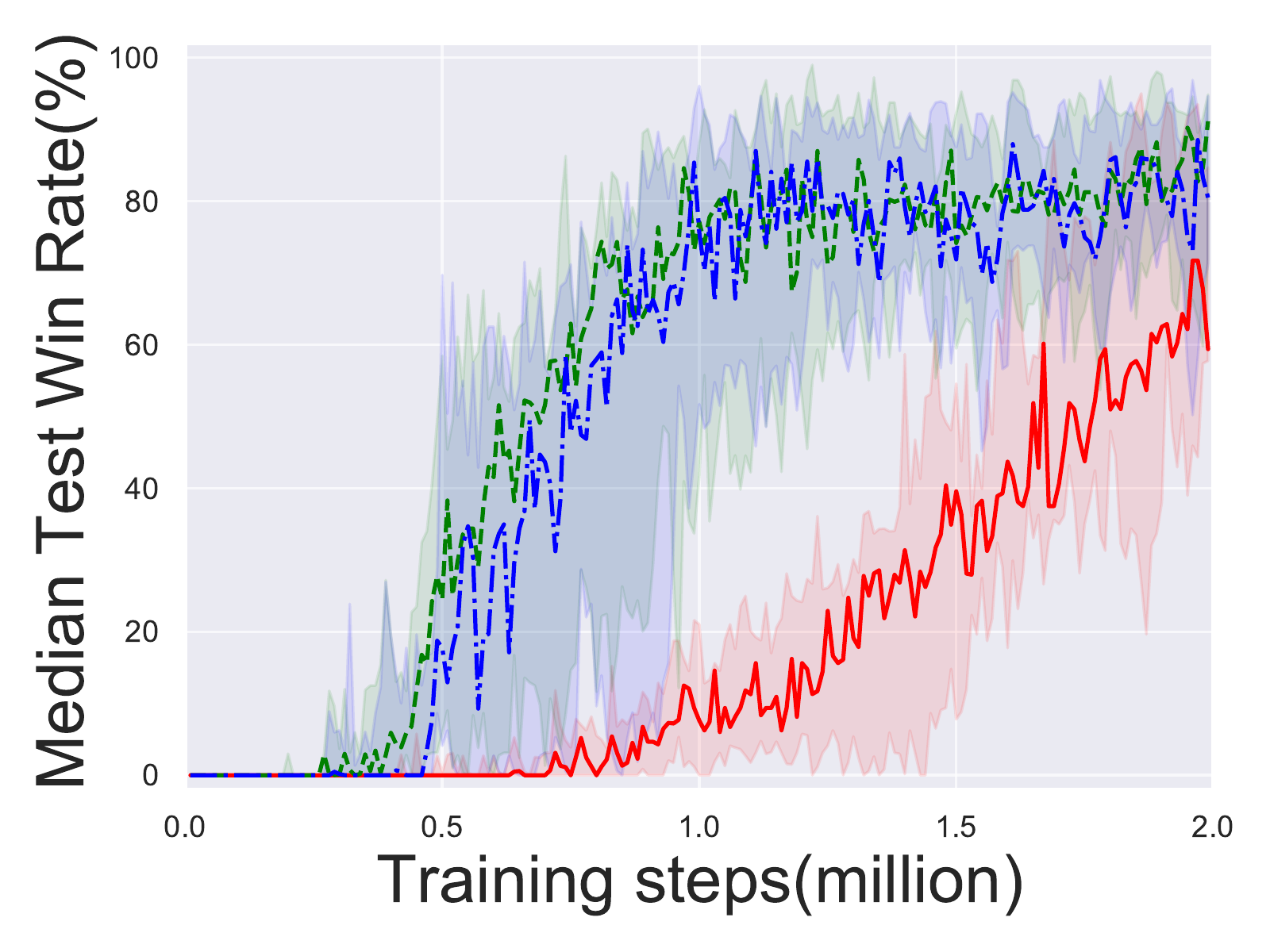}
		\label{QMIX-2c_vs_64zg}
	}
	\hfill
	\subfigure[QPLEX-2c\_vs\_64zg]{
		\includegraphics[width=0.22\textwidth]{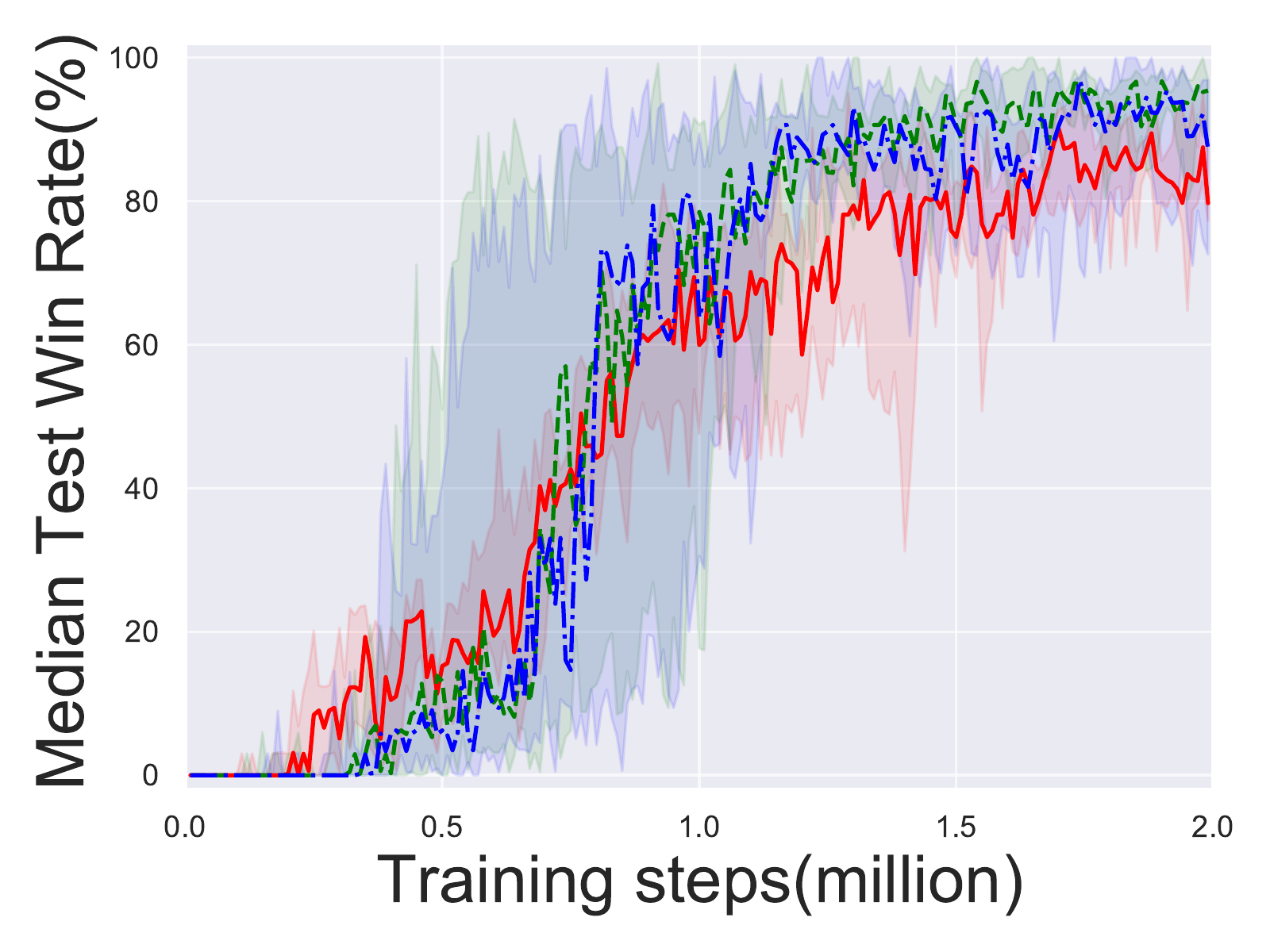}
		\label{QPLEX-2c_vs_64zg}
	}
	\hfill
	\subfigure{
		\includegraphics[width=0.2\textwidth]{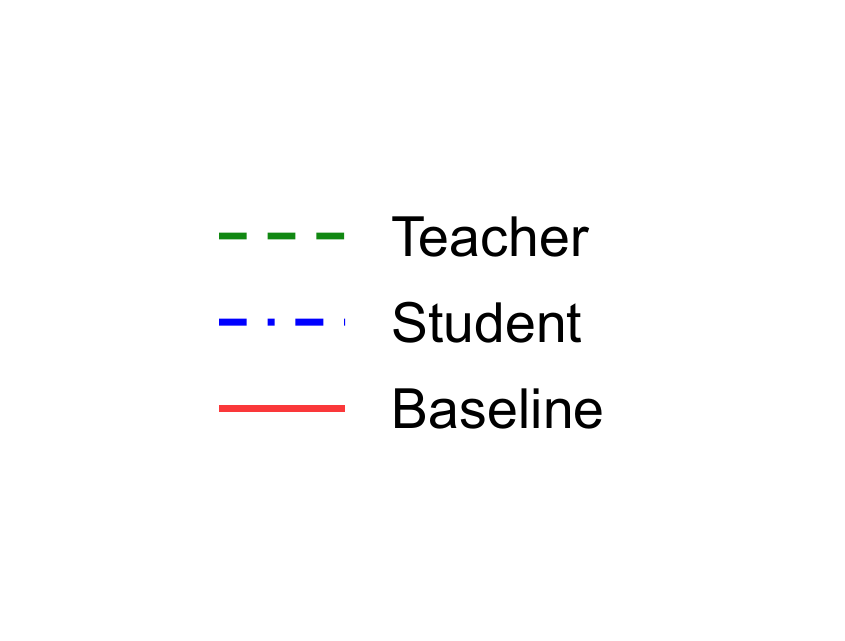}
		\label{legand}
	}
	\caption{
	Test win rates of the base models ($i.e.$, VDN,  QMIX  and  QPLEX)  and the corresponding ones with CTDS algorithm on five different maps of StarCraft II.
	``Baseline'' represents the performance of the original base model; ``Teacher'' and ``Student'' represent the performance of teacher and student module of CTDS, respectively.
	``Baseline'' and ``Student'' achieve decentralized execution.	
	The solid/dashed line shows the median win rate and the shadow area represents the max-min win rate in 5 different random seeds.}
	\label{Experiment}
\end{figure*}

 In the experiment implementation, the different between partial observation and perfect observation depends on the sight range setting. 
 Partial observation restricts the agents from receiving information about allied or enemy units that are out of range, $i.e.$, the features of enemies and allies are all unknown when the corresponding unit beyond the sight range.
For example, the sight range of each agent is set to 2, which means each agent is allowed to only access its own features and other units' features within the distance of 2.
For the teacher module, perfect observation $\hat{o}_i$ is not limited by the sight range, that is to say, each agent can access the full observation of all the agents.
Figure3 shows the examples of the agent with partial and global observations in the SMAC, respectively.
In SMAC, we set partial sight range as 2 and perfect sight range as infinity.
In Combat, partial sight range is 2 and perfect sight range is 4.

\begin{table*}[t]
	\centering
	\resizebox{\textwidth}{!}{
	\renewcommand{\arraystretch}{1.5}
	\begin{tabular}{|c|c|c|c|c|c|c|c|c|c|}%
		\hline
		\multirow{2}{*}{Scenario}  & \multicolumn{3}{|c|}{VDN\cite{sunehag2017value}} & \multicolumn{3}{|c|}{QMIX\cite{rashid2018qmix}} & \multicolumn{3}{|c|}{QPLEX\cite{wang2020qplex}}\\
		\cline{2-10}
		 &Baseline & Teacher & Student & Baseline & Teacher & Student & Baseline & Teacher & Student\\
		\hline
		5m\_vs\_6m & 6.9 $\pm$ 0.2 & 73.8 $\pm$ 0.9 & \textbf{55.0 $\pm$ 0.5} & 21.2 $\pm$ 0.4 & 66.9 $\pm$ 0.5 & \textbf{48.1 $\pm$ 1.0} & 23.8 $\pm$ 1.4 & 62.5 $\pm$ 0.9 & \textbf{56.2 $\pm$ 3.7}\\
		9m\_vs\_11m & 0.6 $\pm$ 0.0 & 9.4 $\pm$ 0.6 & \textbf{12.5 $\pm$ 1.1} & 6.9 $\pm$ 0.1 & 9.4 $\pm$ 0.2 & \textbf{11.2 $\pm$ 0.1} & \textbf{3.1 $\pm$ 0.1} & 0.6 $\pm$ 0.0 & 0.6 $\pm$ 0.0 \\
		27m\_vs\_30m & 0.0 $\pm$ 0.0 & 6.2 $\pm$ 0.1 & \textbf{1.9 $\pm$ 0.0} & 8.8 $\pm$ 1.0 & 58.8 $\pm$ 2.4 & \textbf{42.5 $\pm$ 1.5} & 0.0 $\pm$ 0.0 & 60.9 $\pm$ 5.8 & \textbf{49.2 $\pm$ 8.9}\\
		MMM2 & 0.0 $\pm$ 0.0 & 15.6 $\pm$ 1.2 & \textbf{2.5 $\pm$ 0.1} & 20.6 $\pm$ 1.2 & 80.6 $\pm$ 1.7 & \textbf{28.1 $\pm$ 1.7} & 0.0 $\pm$ 0.0 & 26.6 $\pm$ 0.8 & \textbf{1.6 $\pm$ 0.1}\\
		2c\_vs\_64zg & 44.4 $\pm$ 4.7 & 70.6 $\pm$ 3.6 & \textbf{57.5 $\pm$ 2.6} & 69.4 $\pm$ 2.6 & 82.5 $\pm$ 0.4 & \textbf{80.6$\pm$ 0.7} & 81.9 $\pm$ 0.6 & 93.1 $\pm$ 0.1 & \textbf{87.5 $\pm$ 0.4}\\
		\hline
		Average & 10.4  & 35.1 & \textbf{25.9}  & 25.4 & 59.6 & \textbf{42.1} & 21.8 & 48.7 & \textbf{39.0} \\
		\hline
	\end{tabular}}
	\caption{The performance of the test win rate percentage (including mean and standard deviation) of different base models under different scenarios of StarCraft II.
	``Baseline'' represents the original base model; ``Teacher'' and ``Student'' represent the teacher and student module of CTDS, respectively.
	The one with higher win rate of ``Student'' and ``Baseline'' is highlighted with bold.
	}
	\label{tab:result}
\end{table*}

The algorithms we use are based on the Pymarl algorithm library \cite{samvelyan19smac}.
To reduce randomness, all the reported results are averaged over 5 runs with different seeds.
Training and evaluation schedules such as the testing episode number and training hyper-parameters are kept the same as default hyper-parameters in Pymarl.
Following the base models~\cite{sunehag2017value,rashid2018qmix,wang2020qplex}, we share the parameters of the agent networks across all agents in the teacher/student module to speed up the learning.
Note that the parameters of the teacher module and student module are not shared.
All the parameters in the neural networks are adjusted by a RMSprop optimizer with a learning rate of 0.0005, the same as that used by the baselines.
For all the compared methods, each task is trained for 2 Million steps separately in SMAC and 500k steps in Combat.

\subsection{Performance Evaluation and Discussion}\label{sec:eval}
Our CTDS framework can be applied upon various existing models that follow the CTDE paradigm.
To show the effectiveness of our CTDS framework, we conduct the experiments upon three representative models with different mixing network settings: VDN \cite{sunehag2017value}, QMIX \cite{rashid2018qmix} and QPLEX \cite{wang2020qplex}.


\begin{figure*}[h]
	\centering
	\subfigure[QMIX]{
		\includegraphics[width=0.4\textwidth]{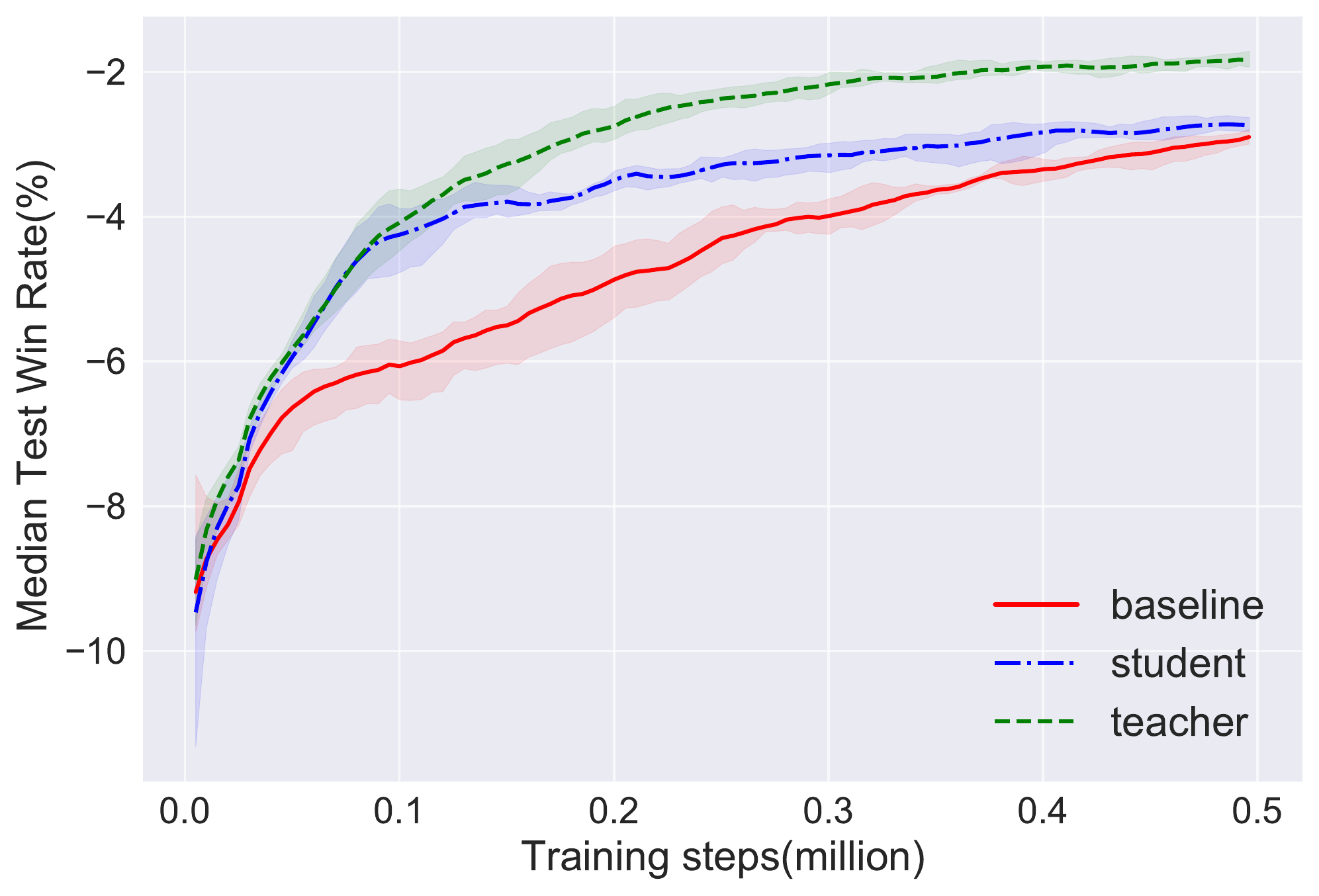}
	}
	\subfigure[QPLEX]{
		\includegraphics[width=0.4\textwidth]{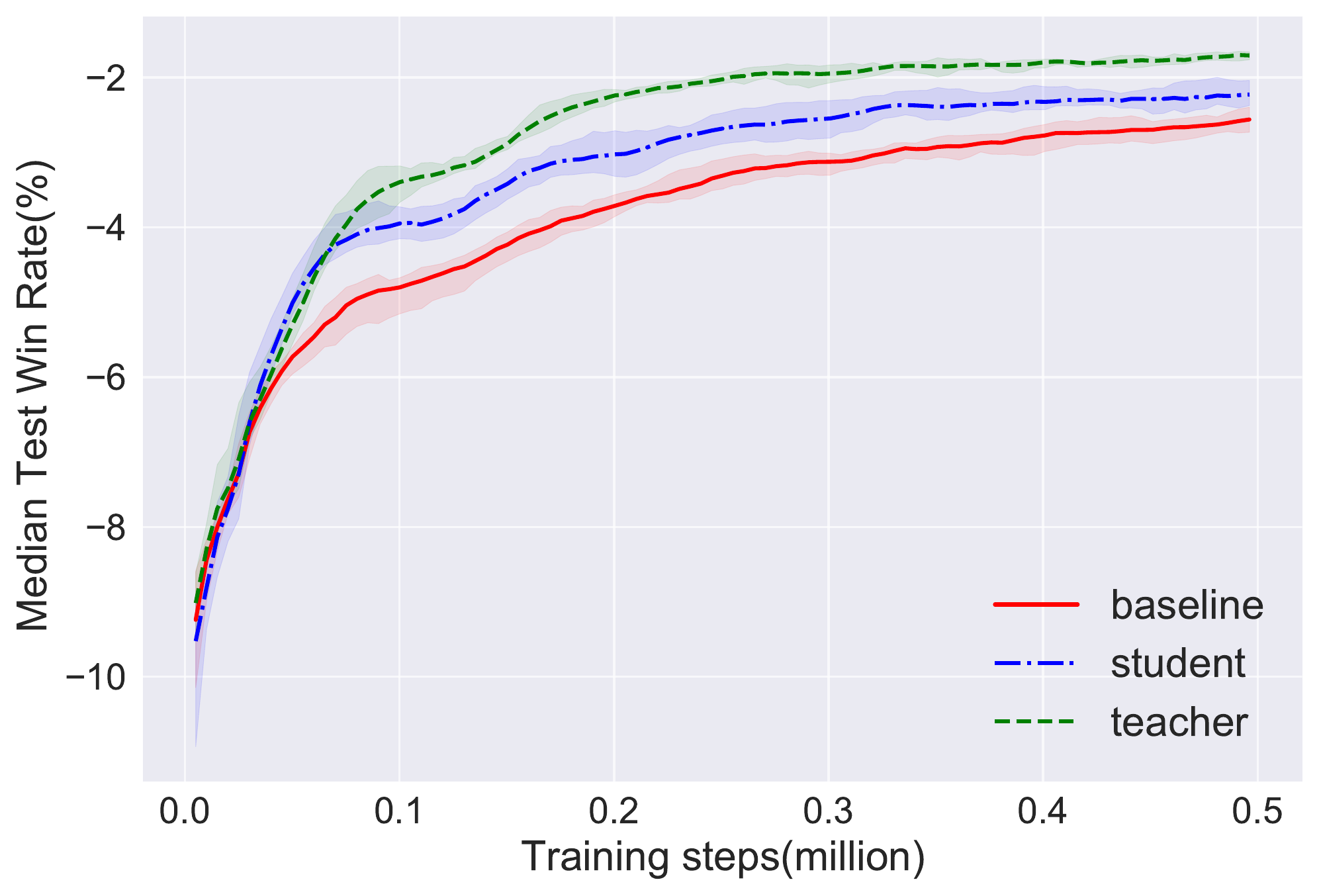}
	}
	\caption{Test win rates of the base models (QMIX and QPLEX) and the corresponding ones with CTDS algorithm in Combat. “Baseline” represents the performance of the original base model; “Teacher” and “Student”represent the performance of teacher and student module of CTDS, respectively. “Baseline” and “Student” achieve decentralized execution. The solid/dashed line shows the median win rate and the shadow area represents the max-min win rate in 5 different random seeds. }
	\label{fig:combat_results}
\end{figure*}

First we conduct the experiment in grid world environment Combat.
The Figure \ref{fig:combat_results} reveals the test win rate curves of CTDS and baselines under the settings of the mixing neural network of QMIX and QPLEX after 500k steps of training.
As demonstrated in Figure \ref{fig:combat_results}, the CTDE paradigm  improves convergence speed at the beginning of trainning and reaches the converging win rate with much lower variance.

The following are the experiment results in StarCraft II.
Table~\ref{tab:result} shows the final performance in terms of the test win rate percentage criteria of different base models under different scenarios.
Here ``Baseline'' means the performance of the original base model which achieves decentralized execution; ``Teacher'' means the performance of the teacher module which relies on the perfect observation thus violates decentralized execution; ``Student'' means the performance of the student module, which distills the knowledge from the teacher and allows decentralized execution.

It can be seen that the teacher always achieves a higher win percentage than the student, which is consistent with our assumption that the full observation brings benefits.
Under the settings of the mixing neural network of VDN, QMIX and QPLEX, after 2 million steps of training, the performance gap between CTDS and the baseline exceeds the average win rate of 15.5\%, 16.7\% and 17.2\%, respectively.
Figure~\ref{Experiment} demonstrates the learning curve of the base model, the teacher module and the student module on different scenarios.
The above results indicate that CTDS framework can boost the performance of the original algorithms to a considerable margin in most scenarios.

\subsection{Sight Range Analysis}

Under the setting of partial observability, agents can only receive the information within the restricted sight range.
All the results reported in Section~\ref{sec:eval} are under the scenario that the sight range is 2.
In this section, we further analyze the generalization of our method with regard to different sight ranges.
\begin{wrapfigure}{r}{0.55\textwidth}
	\centering
    \setlength{\abovecaptionskip}{-0.2cm}
	\includegraphics[width=0.55\textwidth]{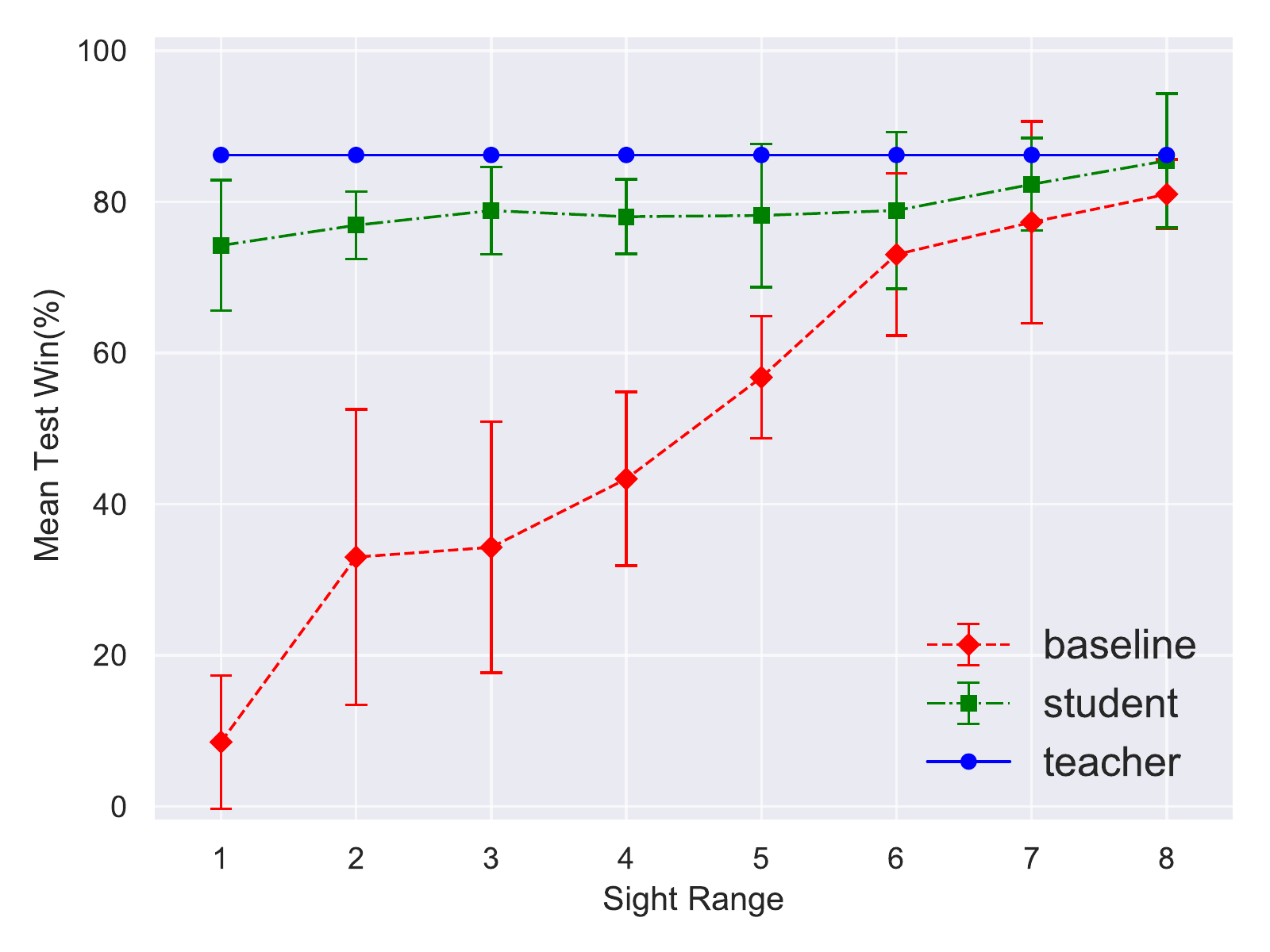}
	\caption{The performance of the test win rate percentage (including mean and standard deviation) of the different sight range. 
	The experiment takes QMIX as the base model on the 2c\_vs\_64zg task after 1 million steps of training. ``Baseline'' represents the performance of the original base model which achieves decentralized execution; ``Teacher'' and ``Student'' represent the performance of teacher and student module of CTDS, respectively.}
	\label{ablation}
\end{wrapfigure}

We adjust the sight range from 1 to 8 in the environment, and investigate the performance of the base model, the teacher module and the student module under each setting, respectively.
Due to the page limit, we only represent the results of methods taking QMIX as the base model on the 2c\_vs\_64zg task of StarCraft II after 1 million steps of training.
As illustrated in Figure~\ref{ablation}, we can see that the performance of the original QMIX significantly increases with the increasing sight range, which indicates that more information about the other agents brings about a higher win percentage.
Besides, the margin gain of the win percentage with one more sight range shows a decreasing trend,
which is consistent with our common sense.
Different from the base model which relies heavily on the sight range, the performance of our CTDS framework is relatively stable under different sight range that increases at a lower margin as the sight range enlarges.
In other words, the performance gap between CTDS and the baseline increases with the decrease of sight range from 8 to 1.
Our CTDS framework thereby has remarkable advantages in the case of the very limited observations.
Overall, the generalization of our framework with the varying of sight range indicates the efficacy of the knowledge distillation mechanism.


\section{Conclusion}
In this paper, we propose a novel cooperative multi-agent reinforcement learning framework, named CTDS, which adopts the teacher module to allocate the team reward during centralized training and employs the student module to allow the agents to take actions only conditioned on local observations for decentralized execution.
To realize decentralized execution, the student module aims to learn the individual Q-values based on only local observations by distilling the knowledge learned by the teacher module.
In this way, CTDS makes full use of the global observation to facilitate the learning process while achieving decentralized execution.
Empirical results demonstrate that CTDS outperforms the recent state-of-the-art baselines both in terms of absolute performance and learning speed, which indicates the efficacy of our method and illustrates the benefits of utilizing centralized observations.
In the future, we will further improve the representations of the full range observations during training to achieve better performance and pay attention to improving efficiency as well.

\bibliographystyle{unsrt}  
\bibliography{references}  






\end{document}